\documentclass[prd,twocolumn,aps,showpacs,nofootinbib,epsfig]{revtex4-1}

\usepackage{amssymb}
\usepackage{amsmath}
\usepackage{graphics}
\usepackage{epsfig}
\usepackage[dvips]{color}

\newcommand{\be}{\begin{equation}}
\newcommand{\ee}{\end{equation}}
\newcommand{\bea}{\begin{eqnarray}}
\newcommand{\eea}{\end{eqnarray}}
\newcommand{\beas}{\begin{eqnarray*}}
\newcommand{\eeas}{\end{eqnarray*}}

\newcommand{\ext}{{\mbox{\tiny{ext}}}}

\begin{document}
\title{Symmetry restoration at finite temperature with weak magnetic
       fields}
\author{Jorge Navarro$^{1,2}$, Angel S\'anchez$^3$, Maria Elena
  Tejeda-Yeomans$^4$, Alejandro Ayala$^1$
and Gabriella Piccinelli$^5$}
\affiliation{$^1$Instituto de Ciencias Nucleares, Universidad Nacional
  Aut\'onoma de
M\'exico, Apartado Postal 70-543, M\'exico Distrito Federal 04510, Mexico.\\
$^2$Departamento de F\'{\i}sica, Universidad del Atl\'antico, Km.~7 antigua
v\'{\i}a a Puerto Colombia, A.A. 1890, Barranquilla, Colombia.\\
$^3$Instituto de F\1sica y Matem\'aticas,
Universidad Michoacana de San Nicol\'as de Hidalgo, Apartado Postal
2-82, Morelia, Michoac\'an 58040, M\'exico.\\
$^4$Departamento de F\'{\i}sica, Universidad de Sonora, Boulevard Luis
Encinas J. y Rosales, Colonia Centro, Hermosillo, Sonora 83000, Mexico. \\
$^5$Centro Tecnol\'ogico, FES Arag\'on, Universidad Nacional Aut\'onoma de
M\'exico, Avenida Rancho Seco S/N, Bosques de Arag\'on, Nezahualc\'oyotl,
Estado de M\'exico 57130, Mexico.}

\begin{abstract}
We study symmetry restoration at finite temperature in the standard model
during the electroweak phase transition in the presence of a weak
magnetic field. We compute the finite temperature effective potential up
to the contribution of ring diagrams, using the broken phase degrees of
freedom, and keep track of the gauge parameter dependence of the results. We
show that under these conditions, the phase transition becomes stronger
first order.
\end{abstract}

\maketitle

\section{Introduction}\label{I}


The properties of the universe's primordial plasma at high temperature play
an important role in the attempts to explain several outstanding questions in
cosmology such as the origin of the matter-antimatter asymmetry. Starting
from symmetric conditions, it was shown by Sakharov~\cite{Sakharov} that
three ingredients are needed to develop a baryon asymmetry: $(i)$ Baryon
number violation processes, $(ii)$ C and CP violation and $(iii)$ departure
from thermal equilibrium. These conditions are met in the Minimal Standard
Model (MSM) during the Electroweak Phase Transition (EWPT) provided that
this be first order. It is however well known that neither the amount of
CP violation nor the strength of the phase transition
are enough to produce and subsequently preserve a possible baryon number at
this stage of the universe evolution~\cite{Gavela, Kajantie1}.

A great deal of effort has been devoted to put forward viable scenarios that
increase the amount of CP violation and/or make the EWPT stronger first
order~\cite{elreview, Pallares}. At the same time, in the
last years, the study of the effects of primordial magnetic fields on other
cosmological processes~\cite{Cheng, Wang, Enqvistnew}, have also been given
attention, including the very same EWPT~\cite{Kajantie, Fiore}.

Since magnetic fields before the EWPT belong to the $U_Y(1)$ group, this kind
of fields properly receive the name of \textit{hypermagnetic}. The
expectation is that in their presence, the order of the EWPT is increased, in
analogy with the case of superconductivity, where an external magnetic field
changes the order of the phase transition from second to first due to the
Meissner effect~\cite{Giovannini,Elmfors,Kajantie}.

The description of the physical processes able to generate magnetic fields
is an old problem in cosmology~\cite{Grasso}. The general approach is to
identify mechanisms for the generation of seed fields~\cite{Barrow,Barrow2}
which can later be amplified into fields on larger scales.
For instance, certain types of inflationary models can produce magnetic
fields extending over horizon distances~\cite{Turner}. Other possibilities
make use of cosmological phase transitions~\cite{phase}
through formation of dipole charge layers on surfaces of phase transition
bubble walls. Although at present there is no conclusive evidence about
the origin of magnetic fields, their existence prior to the EWPT cannot
certainly be ruled out. They have been observed in galaxies, clusters,
intracluster medium and high redshift objects~\cite{31}.

The presence of magnetic fields in the early universe can leave its imprint in
a variety of phenomena. Magnetic fields directly interact with ionized baryons
and indirectly influence photons, through the tight coupling between baryons
and photons at that epoch. Cold dark matter is also indirectly affected,
through gravitational interaction. Thus, a primordial magnetic field imprint
can be searched for (through temperature anisotropies, statistics and
polarization) in the Cosmic Microwave Background (CMB), formation of large
scale structure, gravitational wave background and the nucleosynthesis
processes. A homogeneous
magnetic field would give rise to a dipole anisotropy in the background
radiation. On this basis, Cosmic Background Explorer (COBE) results give an
upper bound on the present equivalent field strength of $B_0\lesssim 10^{-9}$
G~\cite{Barrow,subramanian}.
A lot of work has been done recently in the field of the observational
constraints, considering different aspects of the interaction of primordial
magnetic fields with the CMB, as well as different features and scales of
these cosmic fields (see for example~\cite{Gio} and references therein). In
particular, in a recent work~\cite{Yamazaki2}, using both CMB and large-scale structure observational data, the authors place new constraints on
primordial magnetic fields, both on their magnitude and power spectral
index. This last parameter is important for the discrimination between models
of magnetogenesis. For all cosmological parameters, the same priors as those
adopted in the WMAP analysis are used.  Standard cosmological parameters are
mainly constrained by low multipoles, while primordial magnetic fields
dominate on small angular scales (high multipoles). The degeneracy among
magnetic fields parameters is broken by the different effects of primordial
magnetic fields on the matter power and CMB radiation spectra. An interesting
result of this analysis is that the probability distribution of the parameters
has a maximum for nonzero values of both the magnitude and the power spectral
index. Although these values are still consistent with zero magnetic fields
(thus implying only upper limits on the field strength of $\sim 10^{-9}$ Gauss
on a comoving scale of 1 Mpc) they suggest the possibility of a detection with
forthcoming data.

Recall that the development of the EWPT can be described by means of the MSM
effective potential (EP) where the order parameter is the Higgs vacuum
expectation value $v$~\cite{reviewsEWPT,Petropoulos}. Starting
from a very high temperature,
the minimum of the theory happens for $v=0$, the so called
symmetric phase. Decreasing the temperature produces the EP
to develop a secondary minimum, in case the phase transition is first order.
Since the phase transition does not start until the secondary minimum is
degenerate with the original one, its development in the high temperature
phase can be described in terms of MSM symmetry restored degrees of freedom.
Particle masses depend parametrically on the order parameter and acquire
their fixed values only when the phase transition is completed.

Within the particle physics calculational methods, it is common to work
using the degrees of freedom in the symmetry restored
phase. A drawback of this scheme is that, since the
gauge boson
mass matrix is not diagonal, contributions from gauge bosons to the
EP are harder to compute. Another general aspect of
calculations involving MSM degrees of freedom is that they are usually
performed for a given value of the (covariant) gauge parameter, which makes
it impossible to keep track of a gauge parameter presence, if any, in phase
transition related observables.

In a recent work~\cite{sap}, the development of the EWPT in the MSM from the
symmetric phase has been studied in the presence of a weak external
hypermagnetic field up to the contribution of ring diagrams. The main result
from that work is that the presence of the field strengthens the first order
nature of the phase transition. The result is in agreement with
calculations performed at a classical~\cite{Giovannini} and one-loop
levels~\cite{Elmfors}, as well as with lattice simulations~\cite{Kajantie}. On
the other hand, other analytical non-perturbative approaches where the MSM
finite temperature EP is studied for the case of strong
magnetic fields~\cite{Skalozub1,Skalozub2}, reach the conclusion that these
fields inhibit the first order phase transition. These works attribute the
result to the contribution of light fermion masses which are generally
neglected in other computations. However, they also neglect to consider the
infrared cutoff provided by thermal masses, which casts doubts on their
conclusions. It is thus important to study the phase transition
using the broken phase degrees of freedom, since it will
help to set up the stage for a detailed
analysis of the phase equilibrium conditions as
the EWPT develops.

In this work we generalize the analysis of the development of the EWPT in
the presence of a constant magnetic field, working with the degrees of
freedom in the broken symmetry phase where the symmetry of the theory has been
reduced from $SU(2)_L\times U(1)_Y$ to $U(1)_{\mbox{\small{em}}}$. We work
with an arbitrary value of the (covariant) gauge parameter in the weak field
limit and up to the contribution of the \textit{ring} diagrams, that have
been shown to be crucial for the description of the long wavelength properties
of the theory~\cite{Carrington}. We find that there is a small gauge parameter
dependence on observables such as the critical temperature and the
position of the broken phase minimum, but otherwise confirm that the
presence of a weak magnetic field, increases the order of the phase
transition.

Also, as we will see in the next section, we work explicitly with the
assumption that the hierarchy of scales $eB \ll m^2 \ll T^2$
is obeyed, where we consider $m$ as a generic mass of the problem at the
electroweak scale. Since at the phase transition $m \sim v$, the effective
coupling in the perturbative expansion becomes $g^2T/v$ and thus when $v/T$ is
small ($T/v$ is large), the perturbative series is not reliable.
This issue has been discussed in detail by Arnold and
Espinosa ~\cite{arnold} where they use a power
counting argument whereby the leading temperature terms (after accounting for
the thermal contribution to the effective mass) give rise to corrections
proportional to $g^2T/m$. Notice
however that in this work we do not attempt to claim
that the hierarchy of scales that we worked with makes $v/T$ grow to the extent that both, the sphaleron erasure bound is avoided nor the perturbative
expansion problem is solved. Rather, our purpose is to explore the effective
potential method to see if a hint of a growth of $v/T$ is possible. As we
will see, such growth does happen and in that sense the results are encouraging.
Another recent study of the effective Lagrangian in the MSM using
broken phase degrees of freedom at finite temperature and density can be found
in Ref.~\cite{Erdas2010}.

The paper is organized as follows: In Sec.~\ref{II} we write down the MSM
using the degrees of freedom in the broken symmetry phase. In Sec.~\ref{III}
we lay down the formalism to include weak magnetic fields in the computation
of charged particle propagators. In Sec.~\ref{IV}, we work with these
degrees of freedom to compute particle self-energies that are used in
Sec.~\ref{V} to compute the MSM EP up to the contributions of
ring diagrams. In Sec.~\ref{VI} we study this EP as a
function of the Higgs vacuum expectation value and show that the order of
the EWPT becomes stronger first order in the presence of the magnetic field.
We also study the gauge parameter dependence of the phase transition
parameters. Finally, we conclude and discuss our results in Sec.~\ref{VII} and
leave for the appendices the listing of the one-loop MSM
self-energies in the presence of weak magnetic fields, for arbitrary values of
the gauge parameter.

\section{Minimal Standard Model}\label{II}

Quantization of the MSM in an arbitrary covariant gauge using the
Faddeev-Popov technique involves not only physical fields
such as the gauge bosons $W_{\mu} ,Z_{\mu} ,A_{\mu}$, the neutral Higgs $h$,
and fermions, but also charged and neutral Goldstone bosons $\chi ^{\pm}
,\chi _{3}$, respectively, which are fields that do not propagate
asymptotically. Nevertheless, these last fields appear in loops. Working in a
renormalizable gauge of the t'Hooft type, the so called $R_{\xi}$ gauge, in
order to compensate contributions from the spurious components of the gauge
fields, it is necessary to also introduce Faddeev-Popov ghosts for each of the
gauge  boson fields $\eta_{W}^{\pm},\eta_{Z}$ and $\eta _{\gamma}$. Thus, the
Lagrangian also depends on the choice of the gauge fixing term.

In order to consider all the contributions to the MSM EP,
we write the complete Lagrangian sector by sector, in an arbitrary $R_\xi$
gauge, after the symmetry breaking.
\begin{equation}
   \mathcal{L}= \mathcal{L}_{H}+\mathcal{L}_{gb}
              +\mathcal{L}_{f}+ \mathcal{L}_{Y}
              +\mathcal{L}_{gf}+\mathcal{L}_{FP},
\label{lcompl}
\end{equation}
where $\mathcal{L}_{H},\ \mathcal{L}_{gb},\ \mathcal{L}_{f},\ \mathcal{L}%
_{Y},\ \mathcal{L}_{gf}$ and $\mathcal{L}_{FP}$ are the Lagrangians for the
Higgs, gauge boson, fermion, Yukawa, gauge fixing and Faddeev-Popov sectors,
respectively.

The Lagrangian for the Higgs field is
\begin{equation}
   \mathcal{L}_{H}=(D_{\mu} \Phi)^{\dagger}(D^{\mu} \Phi)+c^2(\Phi^{\dagger}
   \Phi) -\lambda (\Phi^{\dagger} \Phi)^2.
\label{lhiggs}
\end{equation}
The $SU(2)_L\times U(1)_Y$ symmetry of the Lagrangian in Eq.~(\ref{lcompl})
is broken spontaneously by introducing an adequate Higgs scalar $\Phi$,
which develops a vacuum expectation value $v$. The most economical choice
for $\Phi$ is an $SU(2)_L$ doublet of complex fields with hypercharge
($\mbox{\bf \scriptsize{Y}}$) $+1$, given by
\begin{eqnarray}
   \Phi=\frac{1}{\sqrt{2}}\left(
      \begin{array}{rr}
      \chi_2+i\chi_1 &  \\
      v+h-i\chi_3 &
      \end{array}
   \right).
\label{higgsdobl}
\end{eqnarray}
 We
take $h$ as the physical Higgs field with a mass given by
\begin{eqnarray}
   m_H^2 = 3\lambda v^2 -c^2 \equiv m_4^2.
\label{higgsmass}
\end{eqnarray}

From the kinetic energy term on Eq.~(\ref{lhiggs}) one obtains the gauge
boson masses by finding the mass eigenstates in the broken symmetry
phase, which leads to
\begin{equation}
   m_W^2=g^2v^2/4,~~~~  m_Z^2=(g^2+g'^2)v^2/4
\label{mWZvev}
\end{equation}
and the relations between the gauge fields $W^a_{\mu}, B_\mu$ and $
W^\pm_{\mu}, Z_{\mu}, A_{\mu}$ before and after symmetry breaking,
respectively
\begin{eqnarray}
   W^{\pm}_{\mu}&=& \frac{1}{\sqrt{2}}\left(W^{1}_\mu \mp iW^2_{\mu} \right),
   \nonumber \\
   Z_{\mu}&=&\frac{1}{\sqrt{g^2 + g'^2}}\left(g
   W^3_{\mu}-g'B_\mu \right),  \nonumber \\
   A_{\mu}&=&\frac{1}{\sqrt{g^2 + g'^2}}\left(g' W^3_{\mu}
   + g B_\mu \right).  \label{defini}
\end{eqnarray}
Hereafter $g$ and $g'$ are the $SU(2)_L$ and $U(1)_Y$ gauge couplings,
respectively.

In order to obtain the interaction terms between the Higgs field and the
gauge bosons, it is convenient to rewrite the covariant derivative
\begin{equation}
D_{\mu}=\partial_{\mu} +ig\frac{\tau^a}{2}{W^a_{\mu}} + i
\frac{g^{\prime}{\mbox{{\bf \scriptsize{Y}}}}}{2}B^{\prime}_{\mu},
\end{equation}
in terms of the mass eigenstates where $B^{\prime}_\mu=B_\mu+B^\ext_\mu$. This
is explicitly accomplished by applying $D_{\mu}$ in terms of the states defined
in Eq.~(\ref{defini}), to the Higgs field in Eq.~(\ref{higgsdobl}). On the
other hand, the self-interactions of the Higgs field, are obtained from the
scalar potential in Eq.~(\ref{lhiggs}).

The kinetic energy from the $SU(2)_L$ and $U(1)_Y$ gauge bosons is
\begin{eqnarray}
   \mathcal{L}_{gb}=-\frac{1}{4} \mathbf{F}^{\mu \nu} \cdot \mathbf{F}_{\mu
   \nu} - \frac{1}{4} B'^{\mu \nu} B^{\prime}_{\mu \nu},
\label{kingb}
\end{eqnarray}
where
\begin{eqnarray}
   \mathbf{F}_{\mu \nu}&=& \partial_\mu \mathbf{W_\nu}- \partial_\nu
   \mathbf{W}%
   _\mu-g \mathbf{W}_\mu \times \mathbf{W}_\nu  \nonumber \\
   B^{\prime}_{\mu \nu}&=&\partial_\mu B^{\prime}_\nu- \partial_\nu
   B^{\prime}_{\mu}.
\label{fieldstrength}
\end{eqnarray}

After symmetry breaking and introduction of the mass eigenstates, one
obtains from Eq.~(\ref{kingb}) the kinetic term as well as interaction terms
involving three and four gauge fields.

The Lagrangian for the fermion sector is
\begin{eqnarray}
   \mathcal{L}_{f}= & &\overline{\Psi}_R {\not \!\! D}
   \Psi_R + \overline{\Psi}
   _L {\not \!\! D} \Psi_L ,
\label{lfer}
\end{eqnarray}
where $\Psi_{L,R}=\frac{1}{2}(1\mp\gamma_5)\Psi$. Once again, the
fermion-gauge boson interactions are obtained by writing the covariant
derivative in terms of the mass eigenstates after symmetry breaking.

We work in the limit where all fermion masses are negligible,
except the top quark mass. Therefore, the main contribution to the Yukawa
sector is
\begin{eqnarray}
   \mathcal{L}_{Y}= y_t\overline{q_L}{\tilde \Phi}t_R + h.c.,
\label{lyuk}
\end{eqnarray}
where $y_t$ is the top quark Yukawa
coupling, $q_L$ is the third family quark doublet, $t_R$ is the right
handed top quark, ${\tilde \Phi}=i\sigma_2\Phi^*$ and $\sigma_2$ is the second
Pauli matrix. After symmetry breaking this Lagrangian yields for
the top quark mass $m_{t}=\frac{y_t}{\sqrt{2}}v$. To determine
the interaction terms between fermions and the Higgs field, it is necessary
to rotate the fermions to the mass eigenbasis.

In the $R_{\xi}$ gauge, the gauge fixing Lagrangian is
\begin{eqnarray}
   \mathcal{L}_{gf}=&-&\frac{1}{2 \xi} (\partial^{\mu}W_{\mu}^i -\frac{1}{2}
   \xi g v \chi^i)^2  \nonumber \\
   &-&\frac{1}{2 \xi} (\partial^{\mu}Z_{\mu} -\frac{1}{2} \xi g^{\prime}v
   \chi_3)^2  \nonumber \\
   &-&\frac{1}{2 \xi} (\partial^{\mu}A_{\mu})^2,
\label{lgf}
\end{eqnarray}
where $i=1,2$ and $\xi$ is the gauge parameter (which for simplicity we set it
to be the same for all gauge bosons) and $m_i$ stands for the
Goldstone boson masses, given by
\begin{eqnarray}
   m_1^2&=&m^2_2=\lambda v^2-c^2+\xi g^2\frac{v^2}{4}
\label{higssmassess}
\end{eqnarray}
and
\begin{eqnarray}
   m_3^2&=&\lambda v^2-c^2+\xi (g^2+{g^{\prime}}^2)\frac{v^2}{4},
\label{higssmasses}
\end{eqnarray}
respectively.

Since we work with an arbitrary value of the gauge fixing parameter $\xi$,
the ghost fields acquire masses proportional to the corresponding
gauge boson ones $m_{gb}$
\begin{equation}
\label{ghostmasses}
 m_{\eta_{gb}} ^2 = \xi m_{gb}^{2}.
\end{equation}
Notice that these fields do not propagate asymptotically, thus appear only
in internal Feynman diagram lines.

In order to find the corresponding
Faddeev-Popov ghost interactions, the ghost fields are rotated like
the gauge fields in Eq.~(\ref{defini}). We emphasize that by working with an
arbitrary value of $\xi$, the ghost fields
do contribute to the $v$-dependent part of the one-loop EP.
Note that the EP is in principle a gauge dependent object~\cite{Dolan},
however, physical quantities obtained from it should be gauge
independent~\cite{Nielsen}.

\section{Charged particle propagators in the presence of a magnetic
Field}\label{III}

We work with MSM degrees of freedom in the symmetry broken phase, where the
external magnetic field belongs to the $U(1)_{\mbox{\small{em}}}$ group. To
include the effect of the external field, we use Schwinger's proper time
method~\cite{Schwinger}. In the broken phase, we have three kinds of charged
particles that couple to the external field, namely: scalars, fermions and
gauge bosons, whose propagators are
\begin{eqnarray}
   D_{B}(x,x^{\prime}) = \phi(x,x^{\prime}) \int \frac{d^4 k}{(2 \pi)^4}
   e^{-ik \cdot (x-x^{\prime})} D_{B}(k),
\label{scalprop}
\end{eqnarray}
\begin{eqnarray}
   S_{B}(x,x^{\prime}) = \phi(x,x^{\prime}) \int \frac{d^4 k}{(2 \pi)^4}
   e^{-ik \cdot (x-x^{\prime})} S_{B} (k),
\label{ferprop}
\end{eqnarray}
\begin{eqnarray}
   G^{\mu\nu}_{B}(x,x^{\prime})= \phi(x,x^{\prime}) \int \frac{d^4 k}{(2 \pi)^4}
   e^{-i k \cdot (x-x^{\prime})} G^{\mu\nu}_{B} (k),
\label{gaubebosonprop}
\end{eqnarray}
respectively. The phase factor $\phi(x,x^{\prime})$, that breaks translation
invariance, is given by
\begin{eqnarray}
   \phi (x,x^{\prime}) \equiv e^{ie \int^x_{x^{\prime}} d\xi^\mu
   \left[B^{{\mbox{\tiny{ext}}}}_\mu +\frac{1}{2} F_{\mu\nu}
   (\xi-x^{\prime})^{\nu}\right]},  \label{phase}
\end{eqnarray}
where the vector potential $B^{{\mbox{\tiny{ext}}}}_\mu=\frac{B}{2}(0,y,-x,0)
$ gives rise to a constant magnetic field of strength $B$ along the
$\hat{z}$ axis and $F^{{\mbox{\tiny{ext}}}}_{\mu\nu}=\partial_\mu
B^{{\mbox{\tiny{ext}}}}_\nu-\partial_\nu B^{{\mbox{\tiny{ext}}}}_\mu$ is the
external field strength tensor.

The momentum dependent functions $D_{B} (k)$, $S_{B} (k)$ and $G^{\mu\nu}_{B}
(k)$ are given by
\begin{eqnarray}
\label{scalpropmom}
   i D_{B} (k)&=&\int_0^\infty \frac{ds}{\cos{eBs}}  \nonumber \\
   &\times& \exp\left\{ i s (k_{||}^2-k_{\bot}^2 \frac{\tan{eBs}}{eBs}-m^2 +i
   \epsilon)\right\},  \nonumber \\
\end{eqnarray}
\begin{eqnarray}
   i S_{B} (k)&=&\int_0^\infty \frac{ds}{\cos{eBs}}  \nonumber \\
   &\times& \exp\left\{ i s (k_{||}^2-k_{\bot}^2 \frac{\tan{eBs}}{ eBs}-m^2_{f}
   +i \epsilon)\right\}  \nonumber \\
   &\times& \left[ (m_f+{\not \! k}_{||})e^{i eB s \sigma_3} -\frac{{\not \!
   k_\bot}}{\cos{eB s}}\right],  \label{ferpropmom}
\end{eqnarray}
and
\begin{eqnarray}
   iG^{\mu\nu}_{B}(k) &=&\int_0^\infty \frac{ds}{\cos{eBs}}
   e^{ i s(k_{||}^2-k_{\bot}^2 \frac{\tan{eBs}}{eBs})}  \nonumber \\
   &\times& \left\{ e^{-is(m_{gb}^{2}-i\epsilon)} \left[ -g^{\mu\nu}_{||}
   +(e^{2eFs})^{\mu\nu}_{\bot}\right] \right.  \nonumber \\
   &+&\left(\frac{e^{-is(m_{gb}^{2}-i\epsilon)}-
   e^{-is(\xi m_{gb}^{2}-i\epsilon)}}{m_{gb}^{2}}
   \right)  \nonumber \\
   &\times& \left[\left(k^{\mu}+k_{\lambda}F^{\mu\lambda}((\tan(eBs))/B)\right)
   \right.  \nonumber \\
   &\times& \left(k^\nu +k_{\rho}F^{\rho\nu}((\tan(eBs))/B)\right)  \nonumber \\
   &-& \left.\left.i\frac{e}{2}\left(F^{\mu\nu}+
   g_{\bot}^{\mu\nu}B\tan(eBs)\right)\right]\right\}.
\label{bospropmom}
\end{eqnarray}
We use the metric tensor $g^{\mu\nu}= \mbox{diag}(1,-1,-1,-1)$ in which
$g^{\mu\nu}=g^{\mu\nu}_{||}-g_{\bot}^{\mu\nu}$ and the notation
$k_{||}^2=k_0^2-k_3^2$, $k_\perp^2=k_1^2+k_2^2$, and
$\sigma^3=i\gamma^1\gamma^2=-\gamma^5{\not \! u}{\not \! b}$. We also use
\begin{eqnarray}
\label{expperpen}
   (e^{2eFs})^{\mu\nu}_{\bot}&=& g_{\bot}^{\mu\nu}\cos (2eBs)
   -F^{\mu\nu}\frac{\sin (eBs)}{B}
\end{eqnarray}
where throughout $u^\mu$ and $b^\mu$ are four-vectors describing the plasma
rest frame and the direction of the magnetic field, respectively. In the rest
frame, these are given by
\begin{equation}
   u=(1,0,0,0), ~~~~b=(0,0,0,1).
\label{uyb}
\end{equation}

It has been shown that, by deforming the contour of integration,
Eqs.~(\ref{scalpropmom}) and~(\ref{ferpropmom}) can be written
as~\cite{nosotros,Tzuu}
\begin{eqnarray}
   iD_{B}(k)=2i\sum_{l=0}^{\infty}\frac{(-1)^lL_l(\frac{2k_\perp^2}{e B})
   {\mathrm e}^{-\frac{k^2_\perp}{e B}}}{k^2_{||}-(2l+1)e B-m^2+i\epsilon},
\label{scalpropsum}
\end{eqnarray}
\begin{eqnarray}
   iS_{B}(k)= i \sum^\infty_{l=0} \frac{d_l(\frac{k_\perp^2}{e B})D +
   d^{\prime}_l(\frac{k_\perp^2}{e B}) \bar D}{k^2_{||}-2 l e B-m_f^2 +
   i\epsilon} + \frac{{\not \! k_{\bot}}}{k^2_\perp},  \label{ferpropsum}
\end{eqnarray}
where $d_l(\alpha)\equiv (-1)^n e^{-\alpha} L^{-1}_l(2\alpha)$,
$d^{\prime}_n=\partial d_n/\partial \alpha$,
\begin{eqnarray}
   D &=& (m_f+{\not \! k_{||}})+ {\not \! k_{\perp}} \frac{m_f^2-k^2_{||}}{ {
   k^2_{\perp}}},  \nonumber \\
   \bar D &=& \gamma_5 {\not \! u}{\not \! b}(m_f + {\not \! k_{||}}),
\label{DDe}
\end{eqnarray}
and $L_l$, $L_l^m$ are Laguerre and Associated Laguerre polynomials,
respectively. It can also be shown that by performing a similar analysis as in
Refs.~\cite{nosotros,Tzuu}, Eq.~(\ref{bospropmom}) can be written as
~\cite{procNos}

\begin{eqnarray}
\label{bospropsum}
   iG_{B}^{\mu\nu}(k)&=& \sum_{\lambda =-1}^{1}\sum_{l=0}^{\infty}
   \frac{2i (-1)^l {\mathrm e}^{-\frac{k^2_\perp}{e B}}}
   {k^2_{||}-(2l+2\lambda+1)e B-m_{gb}^{2}+i\epsilon} \nonumber \\
   && \hspace{-1.5cm} \times \Biggl[ T^{\mu\nu } +\frac{
   m_{gb}^{2}(1-\xi)P^{\mu\nu}}{k_{\parallel}^{2}-(2l+2\lambda+1) eB-\xi
   m_{gb}^{2}+i\epsilon} \Biggr]
   \nonumber \\
   && \hspace{-1.5cm} \times
   L_l\Biggr(\frac{2k_\perp^2}{e B}\Biggl)
\end{eqnarray}
where
\begin{eqnarray}
   T^{\mu\nu }&=&\sum_{\lambda =-1}^{1}\left[g^{\mu \nu} \left(\left|
   \lambda\right| -1\right)\right.  \nonumber \\
   &-&2\left. g_{\perp}^{\mu\nu}\frac{\left(\left| 3\lambda \right| -2\right)}
   {\left(2i\right)^{2}}+2\frac{F^{\mu \nu }}{B} \left( \frac{\lambda }{4i}
   \right)\right],  \nonumber \\
   P^{\mu \nu}&=&\frac{1}{m_{gb}^{2}}[(k^{\mu}k^{\nu}-i\frac{e}{2}F^{\mu \nu})
   +(\frac{k_\lambda k^\nu F^{\mu\lambda} }{B}  \nonumber \\
   &+&\frac{k_{\rho }k^{\mu }F^{\rho\nu}}{B} +i\frac{eB}{2}
   g_{\perp }^{\mu \nu})(ieB)\frac{d}{dk^2_\perp}  \nonumber \\
   &+&(\frac{k_{\lambda }k_{\rho }F^{\mu \lambda } F^{\rho\nu  }}{B^{2}})
   (ieB)^2\frac{d^2}{d(k^2_\perp)^2}].
\label{defbospropsum}
\end{eqnarray}

In order to determine the appropriate order of energy scales during the
development of the the EWPT, one resorts to bounds on the strength of the
magnetic fields imposed by cosmological processes in the early universe.
The relation between the strength of large scale magnetic fields and
temperature is obtained from the requirement that the magnetic energy
density $\rho_{mag} \sim B^2$ should be smaller than the overall radiation
energy density $\rho_{rad} \sim T^4$ at nucleosynthesis, in order to
preserve the estimated abundances of light elements. With this, one obtains
the simple bound $B\lesssim T^2$~\cite{Maartens}. Furthermore, to guarantee
stability conditions against the formation of a $W$-condensate~\cite{Ambjorn},
one obtains that the field is also weak compared to $m_W^2$.
We want to study the EWPT transition via the analysis of the
EP when its minima change in the presence of magnetic fields,
we will work explicitly with the assumption that the hierarchy of scales
\begin{eqnarray}
   eB \ll m^2 \ll T^2,
\label{hierarchy}
\end{eqnarray}
is obeyed, where we consider $m$ as a generic mass of the problem at the
electroweak scale. It is important to address the same problem with the
hierarchy of the mass and the magnetic field scales switched and this
will be done elsewhere~\cite{AyalaReverse}. We can thus perform a weak
field expansion in
Eqs.~(\ref{scalpropsum}),~(\ref{ferpropsum}) and~(\ref{bospropsum}) which
allows us to carry out the summation over Landau levels to write the scalar,
fermion and gauge boson propagators as power series in $eB$, that up to order
$(eB)^2$ read as~\cite{nosotros,Tzuu,procNos}
\begin{eqnarray}
\label{scalpropweak}
   D_{B}(k)=\frac{1}{k^2-m^2}\left( 1-\frac{(eB)^2}{(k^2-m^2)^2}- \frac{2(eB)^2
   k_\perp^2}{(k^2-m^2)^3}\right),
   \nonumber \\
\end{eqnarray}
\begin{eqnarray}
   {S_{B}(k)}= \frac{{\not \! k}+m_f}{{\not \! k}^2-m_f^2}+
   \frac{\gamma_5 {\not \! u}{\not \! b}(k_{||}+m_f)(eB)}{(k^2-m_f^2)^3}
   \nonumber \\
   -\frac{2(eB)^2 k_\perp^2}{(k^2-m_f^2)^4} (m_f+{\not \! k_{||}}+{\not \!
   k_\perp}\frac{m_f^2-k_{||}^2}{k_\perp^2}),
\label{ferpropweak}
\end{eqnarray}
and
\begin{widetext}
\bea
   G^{\mu\nu}_{B}(k)&=&-i \left(\frac{g^{\mu\nu}-(1-\xi)\frac{k^{\mu}k^{\nu}}
   {k^2-\xi m_{gb}^{2}}}{k^2-m_{gb}^{2}}\right)-
   \left(eB\right) \left[\frac{k_{\rho}}{m_{gb}^{2}
   }\left(  k^{\nu}\frac{F^{\mu\rho}}{B}+k^{\mu}\frac{F^{\rho\nu}}
   {B}\right)\right.\bigskip
   \nonumber\\
   &\times&\left.\left(\frac{1}{\left(k^{2}-m_{gb}^{2}\right)^{2}}-
   \frac{1}{\left(k^{2}-\xi m^{2}_{gb}\right)^{2}}\right)-
   \frac{F^{\mu\nu}}{B}\left(\frac {2}{\left(k^{2}-m^{2}_{gb}\right)^{2}}+
   \frac{\left(1-\xi\right)}{2\left(
   k^{2}-m^{2}_{gb}\right) \left(k^{2}-\xi m_{gb}^{2}\right)}\right)\right]
   \nonumber\\
   &+&i\left(eB\right)^{2} \left[\frac{g^{\mu\nu}+4g_{\bot}^{\mu\nu}}{\left(
   k^{2}-m_{gb}^{2}\right)^{3}}+\frac{2g^{\mu\nu}k_{\bot}^{2}}{\left(
   k^{2}-m_{gb}^{2}\right)^{4}}-\frac{k^{\mu}k^{\nu}}{m_{gb}^{2}}\left(  \frac
   {1}{\left(  k^{2}-m_{gb}^{2}\right)  ^{3}}-\frac{1}{\left(k^{2}-\xi m_{gb}
   ^{2}\right)^{3}}\right)\right.\bigskip
   \nonumber\\
   &-&2\frac{k^{\mu}k^{\nu}}{m_{gb}^{2}}k_{\bot}^{2}\left(  \frac{1}{\left(
   k^{2}-m_{gb}^{2}\right)  ^{4}}-\frac{1}{\left(  k^{2}-\xi m^{2}_{gb}\right)^{4}
   }\right)  +\frac{g_{\perp}^{\mu\nu}}{2m^{2}_{gb}}\left(  \frac{1}{\left(
   k^{2}-m_{gb}^{2}\right)  ^{2}}-\frac{1}{\left(  k^{2}-\xi m^{2}_{gb}\right)^{2}
   }\right)
   \nonumber\\
   &-&\left.\frac{2}{m_{gb}^{2}}\left(
   \frac{k_{\lambda}F^{\mu\lambda}k_{\rho}
   F^{\rho\nu}}{B^{2}}\right)  \left(  \frac{1}{\left(  k^{2}-m^{2}_{gb}\right)
   ^{3}
   }-\frac{1}{\left(k^{2}-\xi m^{2}_{gb}\right)^{3}}\right)\right]
\label{bospropweak}
\eea
\end{widetext}
respectively. Notice that for the assumed hierarchy of energy scales to be
valid, one should stay away from working with the value for the gauge parameter $\xi=0$, since, according to Eq.~(\ref{ghostmasses}) this choice would lead to the vanishing of the ghosts masses.

\section{Self-energies}\label{IV}

In this section we compute the MSM self-energies that are in turn used for
the computation of the ring diagrams in the EP.

It is well known that in the absence of an external magnetic field, the MSM
thermal self-energies are gauge independent when considering only the
leading contributions in temperature~\cite{LeBellac}. However, as we will
show, when considering the effects of a weak external magnetic field, these
self-energies turn out to be gauge dependent.

In what follows, we work in the imaginary-time formalism of thermal field
theory. First, we note that the integration over four-momenta is carried out
in Euclidean space with $k_0=ik_4$, this means that
\begin{eqnarray}
   \int \frac{d^4k}{(2\pi)^4} \rightarrow i \int \frac{d^4k_E}{(2\pi)^4}.
\label{minkeuc}
\end{eqnarray}
Next, we recall that boson energies take on discrete values, namely
$k_4=\omega_n=2n \pi T$ with $n$ an integer, and thus
\begin{eqnarray}
   \int \frac{d^4k_E}{(2\pi)^4} \rightarrow T \sum_n \int \frac{d^3k}{(2\pi)^3}.
\label{sumtoint}
\end{eqnarray}

\subsection{Higgs boson}\label{IVa}

\begin{figure}[t!] 
{\centering
{
\epsfig{file=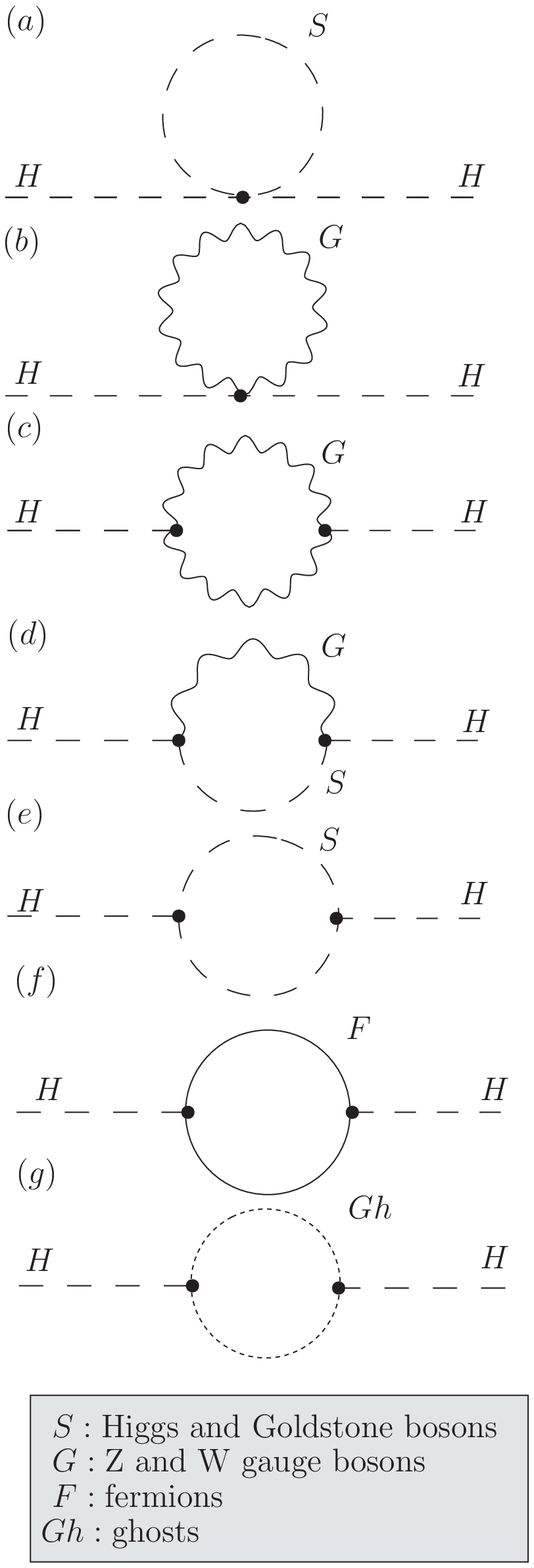, width=0.9\columnwidth}
}
\par}
\caption{Feynman diagrams for the Higgs boson self-energy.}
\label{fig1}
\end{figure}

Figure~\ref{fig1} shows the diagrams that contribute to the Higgs boson
self-energies affected by the magnetic field. Let us explicitly compute the
momentum independent diagram shown in Fig.~\ref{fig1}($a$) for a single
scalar field running in the loop, $\Pi_a^{H-S}$. In the weak field limit, its expression is
\begin{eqnarray}
   \Pi^{H-S}_{a}=\frac{\lambda}{4} T \sum_n \int \frac{d^3 \mathbf{k}}
   {(2\pi)^3}
D_{B}(\omega_n,\mathbf{k};m^2
),
\label{piselfcons}
\end{eqnarray}
where $D_{B}$ is given by Eq.~(\ref{scalpropweak}).

Using the Euclidean version of Eq.~(\ref{scalpropweak}), we have
\begin{eqnarray}
   \Pi^{H-S}_{a}&=& 2{\lambda} T \sum_n \int \frac{d^3 \mathbf{k}}
   {(2\pi)^3}\frac{1}{(\omega_n^2+\mathbf{k}^2+
m^2)}
   \nonumber \\
   &\times& \left(1 - \frac{(eB)^2}{(\omega_n^2+\mathbf{k}^2+
m^2)^2}
   \right.  \nonumber \\
   &+&\left.\frac{2\mathbf{k}_\perp^{2}(eB)^2}{(\omega_n^2+\mathbf{k}^2+
m^2)^3} \right).
\label{autohiggsaaa}
\end{eqnarray}
The integrand
in Eq.~(\ref{autohiggsaaa}) contains terms whose general form is
\begin{eqnarray}
   I_{\alpha}(\mathbf{k};m^2)\equiv\frac{1}{[\omega_n^2+ \mathbf{k}
   ^2+m^2]^\alpha}.
\label{iabbb}
\end{eqnarray}
The integral involving the sum over Matsubara frequencies is performed by
resorting to Refs.~\cite{Bedingham,mishaps}. For the terms with $n\neq 0$ the
result
is
\bea
   T\sum_{n} \int \frac{d^d k}{(2\pi)^d}
   {\bf k}^{2a}{\omega^{2t}_{n}}I_{\alpha}({\bf k};m^2) &=&
   \frac{(2T) }{(4\pi)^{d/2}\Gamma(\alpha)} \nonumber \\
&&\hspace{-3.5cm} \times  (2\pi T)^{d+2a+2t-2\alpha}
  \frac{\Gamma(\frac{d}{2}+a)}{\Gamma(\frac{d}{2})}\mu^{2\epsilon}
\nonumber \\
&&\hspace{-3.5cm}\times    \sum_{j=0}^\infty \frac{(-1)^j}{j!}
   \zeta(2(j+\alpha-t-\frac{d}{2}-a))
   \nonumber \\
   &&\hspace{-3.5cm}\times  \Gamma(j+\alpha-\frac{d}{2}-a)
   \left(\frac{m}{2\pi T}\right)^{2j},
\label{highT}
\eea
whereas for the term $n=0$ we get
\bea
   T \int \frac{d^d k}{(2\pi)^d}\frac{{\bf k}^{2a}}{({\bf
   k}^{2}+m^2)^\alpha} &=&
   \frac{T}{(4\pi)^{d/2}}
   \frac{\Gamma(\frac{d}{2}+a)}{\Gamma(\frac{d}{2})} \nonumber \\
&& \hspace{-2.5cm}\times  \frac{\Gamma(\alpha -\frac{d}{2}-a)}{\Gamma(\alpha)}
   \left(\frac{1}{m^{2}} \right)^{\alpha -\frac{d}{2}-a},
\label{highT00}
\eea
where $\zeta$ is the modified Riemann Zeta function, $\mu$ is the energy
scale of dimensional regularization and $d=3-2\epsilon$.

In terms of Eq.~(\ref{iabbb}), Eq.~(\ref{autohiggsaaa}) is written as
\begin{eqnarray}
   \Pi^{H-S}_{a}&=& 2{\lambda} T \sum_{n} \int \frac{d^3 \mathbf{k}}{
   (2\pi)^3} \Big[I_{1}(\mathbf{k};m_i^2)
   \nonumber \\
   &-&\frac{(eB)^2}{3}\Big(I_{3}(\mathbf{k};m_i^2)- 2 \mathbf{k}^2
   I_{4}(\mathbf{k};m_i^2)
   \Big)\Big].  \label{autohiggsA}
\end{eqnarray}
Using Eq.~(\ref{highT}) for the terms with $n\neq 0$ and Eq.~(\ref{highT00})
for $n=0$, and after considering the contributions from all scalar fields
running in the loop, we get
\begin{eqnarray}
   \Pi^{H}_{a}&=&
\lambda \frac{T^2  }{2}
\left[ 1 - \frac{1}{2\pi T} \left(2 m_1 + m_3 + 3 m_4 \right)\right]
\nonumber \\
&&-\lambda\frac{(eB)^2 }{48 \pi m_1^2} \left[ \frac{T}{m_1}
+\frac{\zeta(3) m_1^2}{4 \pi^3 T^2} \right],
\label{autohiggsaa}
\end{eqnarray}
where
$m_i$ stands for the Goldstone boson masses, given by Eq.~(\ref{higssmassess}).

The contribution from the diagram in Fig.~\ref{fig1}($b$) for a $W$ boson
running in the loop is
\begin{eqnarray}
   \Pi^{H-W}_{b}=\frac{g^2}{2} T \sum_n \int \frac{d^3 \mathbf{k}}{(2\pi)^3}
   G^{\mu}_{B\;\mu}(\omega_n,\mathbf{k};m_W).
\label{piselfbos}
\end{eqnarray}

We proceed to compute Eq.~(\ref{piselfbos}), as well as the rest of the
self-energies, in the infrared limit~\cite{weldon}
$q_0=0,\mathbf{q}\rightarrow 0$, unless explicitly stated.

In terms of the function $I_\alpha$ defined in
Eq.~(\ref{iabbb}) we can write Eq.~(\ref{piselfbos}) as
\begin{eqnarray}
\Pi^{H-W}_{b}&=&
\frac{g^2}{2m_W^2} T \sum_{n} \int \frac{d^3\mathbf{k}}
   {(2\pi)^3} \Big\{
  4m_W^2 I_{1}(\mathbf{k};m_W^2)
   \nonumber \\
   &+& (\omega_ n^2+\mathbf{k}^2) \left[ I_{1}(\mathbf{k};m_W^2)
   - I_{1}(\mathbf{k};\xi m_W^2) \right] \nonumber \\
   &+&\frac{(eB)^2}{3}\Big[
   -6(I_{2}(\mathbf{k};m_W^2)-I_{2}(\mathbf{k};\xi m_W^2))
   \nonumber \\
   &-&(3\omega _n^2+7\mathbf{k}^2)(I_{3}(\mathbf{k};m_W^2)
   -I_{3}(\mathbf{k},\xi m_W^2))
   \nonumber \\
   &+&  4\mathbf{k}^2 (\omega_n^2 + \mathbf{k}^2) (I_{4}(\mathbf{k};m_W^2) -
   I_{4}(\mathbf{k};\xi m_W^2))
   \nonumber \\
  &+&36 m_W^2 I_{3}(\mathbf{k};m_W^2) + 16 \mathbf{k}^2 m_W^2
   I_{4}(\mathbf{k};m_W^2)
   \Big]
   \Big\}. \nonumber \\
\label{autohiggsb}
\end{eqnarray}
Using Eq.~(\ref{highT}) for the terms with $n \neq 0$ and Eq.~(\ref{highT00})
for the terms with $n=0$ and including all contributions from the gauge bosons
in the loop, we get
\begin{eqnarray}
   \Pi^{H}_{b}&=&
\frac{T^2}{16} \left[g^2(3+\xi)+g'^2\left(1+\frac{\xi }{3} \right)
  \right. \nonumber \\
&&\left. - \frac{3 + \xi ^{3/2}}{\pi T} \left(2 g^2 m_W
+ (g^2 + g'^2) m_Z \right)\right] \nonumber \\
&&  + \frac{(eB)^2 g^2}{64 \pi  m_W^2} \left[
\left(3+\frac{35}{3 \xi^{1/2}}\right) \frac{T}{m_W}+\frac{11 \zeta(3)
  m_W^2}{3 \pi ^3 T^2} \right].   \nonumber \\
\label{autohiggsbb}
\end{eqnarray}
In a similar fashion the contributions from the diagrams ($c$) to
($g$) in Fig.~\ref{fig1}, in the infrared limit, can be computed and
are given Appendix A for an arbitrary value of the gauge parameter
$\xi$.

The leading contribution to the Higgs self-energy is obtained by adding
expressions (a)-(g) in Appendix A, keeping the leading term for each
contribution. The result is gauge parameter independent and is explicitly
given by
\begin{eqnarray}
   \Pi_1&=& \frac{T^2}{4} \left\{ \frac{3}{4}g^2+\frac{1}{4}g'^2+2
   \lambda+f^2 \right\}.
\label{pi1}
\end{eqnarray}
Notice that for large values
of the top quark mass, namely a large coupling constant $f$, a perturbative
calculation is not entirely
justified. Nevertheless, here we consider our calculation as an analytical
tool to explore this non-perturbative domain.

In expressions (a)-(g) for the Higgs self-energy in
Appendix A, we have kept terms representing the leading contribution of each
kind arising in the calculation, namely, terms that after extracting a $T^2$
factor, are of order $(eB)^2/T^4$, $m/T$, $v^2/Tm$
and $(eB)^2/Tm^3$, where $m$ is a generic mass. For the hierarchy of scales
considered, the first kind of terms can be safely
neglected. Also,
terms of order $m/T$ are small. These kind of terms include
ratios of gauge boson masses to the temperature, which according to
Eq.~(\ref{mWZvev}) leads to ratios of $v/T$. When the
masses appear in the denominator and are the scalar ones, the terms are
potentially dangerous since their square can become negative. However, as we
will soon show, these kind of terms are
naturally canceled in the effective potential and substituted by
terms where the mass that contributes is the thermal one,
defined by
\bea
   \tilde{m}_i=\sqrt{m_i^2 + \Pi_1}.
   \label{scalarthermalmass}
\eea
When the gauge boson masses appear in the denominators throughout, one should
recall that such masses are to be thought of as being computed in the
broken phase. However, since our analysis considers the effective potential as
a function of $v\geq 0$, we will implement such restriction by replacing in the
effective potential the gauge boson masses appearing in denominators by the
thermal ones, whose definition, in analogy to the scalar case, is given as the
square root of the sum of the gauge boson mass squared and the
leading term for the corresponding self-energy, to be defined shortly [see
Eqs.~(\ref{massTZF}) and~(\ref{PiTZF1})].

\subsection{Gauge bosons}

To express the gauge boson self-energies, in the presence of the external
field, we notice that we have three independent vectors to our disposal to
form tensor structures transverse to the gauge boson momentum $q^\mu$,
namely $q^\mu$, $u^\mu$ and $b^\mu$, where these last two
vectors are given in the rest frame by Eq.~(\ref{uyb}). This means
that in general, these
self-energies can be written as linear combinations of nine independent
structures~\cite{Dolivo, RingQED}. Since we are interested in considering the
infrared limit, $q_0=0,\mathbf{q}\rightarrow0$, only $u^\mu$ and $b^\mu$
remain. Notice that the correct symmetry property for the self-energy is
$\Pi^{\mu\nu}(q)=\Pi^{\nu\mu}(-q)$~\cite{nievespal}. However, in
the infrared limit, this condition means that the self-energy must be
symmetric under the exchange of the Lorentz indices and therefore we can
write:
\begin{eqnarray}
   \Pi^{\mu \nu}= \Pi^Q Q^{\mu\nu} + \Pi^R R^{\mu\nu} +
   \Pi^S S^{\mu\nu} + \Pi^M g^{\mu\nu},
\label{pibosonu1}
\end{eqnarray}
where
\begin{eqnarray}
   Q^{\mu\nu}&=&u^\mu u^\nu,  \nonumber \\
   R^{\mu\nu}&=&b^\mu b^\nu,  \nonumber \\
   S^{\mu\nu}&=&u^\mu b^\nu + u^\nu b^\mu,
\label{base}
\end{eqnarray}
and the transversality condition $q_\mu\Pi^{\mu\nu}=0$ is trivially
satisfied in the infrared limit. Also, note that working in the rest
frame of the medium [see Eq.~(\ref{uyb})],
\begin{eqnarray}
   \Pi^{00}=\Pi^{Q}+\Pi^{M}.
\label{PiQM}
\end{eqnarray}

\begin{figure}[t!] 
{\centering
{
\epsfig{file=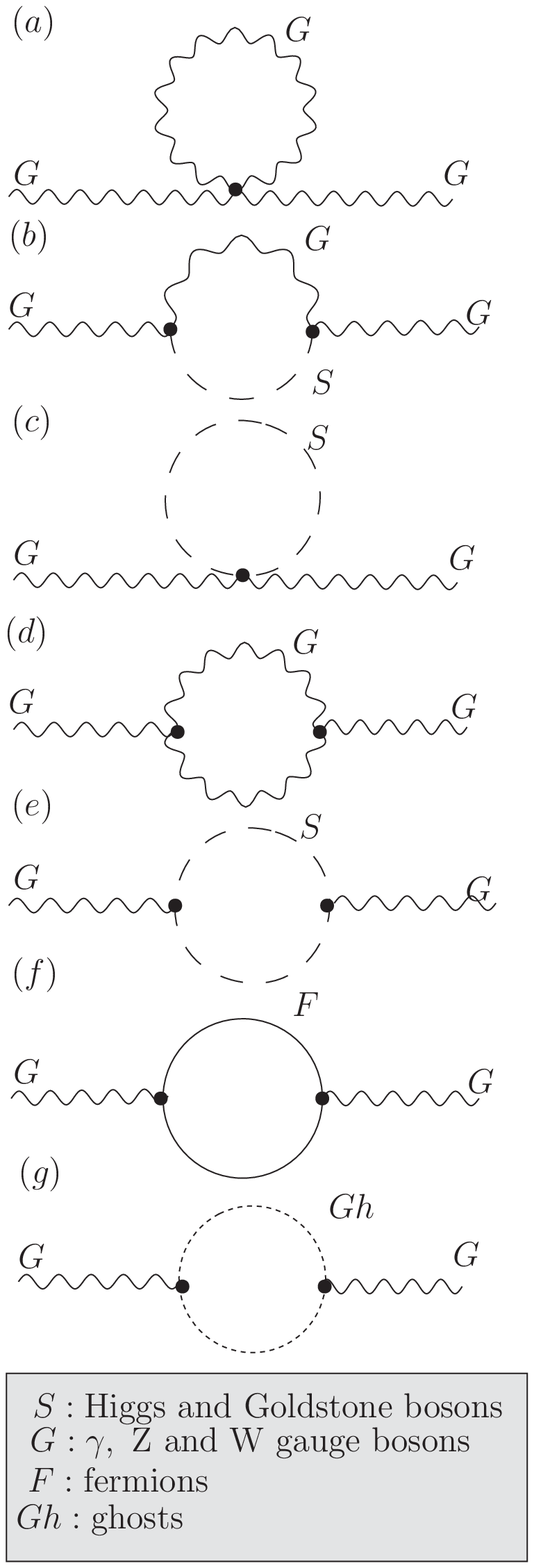, width=0.9\columnwidth}
}
\par}
\caption{Feynman diagrams for the Gauge bosons self-energies.}
\label{fig2}
\end{figure}

Figure~\ref{fig2} shows the gauge boson self-energy diagrams,
which involve Goldstone bosons, fermions as well
as gauge bosons in the loop. Let us
explicitly compute the component $\Pi^{00}_a$ for the self-energy
diagram shown in Fig.~\ref{fig2}(a) for the case
where the external particle and the loop particle are $W$-bosons. The
explicit expression for all the components of the tensor is
\begin{eqnarray}
   \Pi^{\mu\nu\;W}_{(a)}(q)&=&ig^2(2g^{\mu\rho}g^{\nu\sigma} -
   g^{\mu\nu}g^{\rho\sigma} - g^{\mu\sigma}g^{\rho\nu})
   \nonumber \\
   &\times&\int \frac{d^4 k}{(2\pi)^4} G_{B\;\rho\sigma}(k).
\label{auto1a}
\end{eqnarray}
Using the Euclidean version of the gauge boson propagator $G^{\mu\nu}_{B}$
obtained from Eq.~(\ref{bospropweak}), and in terms of the function
$I_{\alpha}$ defined in Eq.~(\ref{iabbb}), we can write the component that is being considered at finite temperature as
\bea
   \Pi^{W-W}_{a} &=& \frac{g^2}{3m_W^2}
   T \sum_n \int \frac{d^3 \mathbf{k}}{(2\pi)^3}
   \Big\{
   9m_W^2I_{1}({\bf k};m_W^2) \nonumber \\
   &+&3\mathbf{k}^2(I_{1}({\bf k};m_W^2) -I_{1}({\bf k};\xi
   m_W^2) ) \nonumber\\
   &+&(eB)^2\left[
   - 3\Big(I_{2}({\bf k};m_W^2)-I_{2}({\bf k},0;\xi
   m_W^2)\Big) \right. \nonumber \\
   &-& \left. 7{\bf k}^{2}\Big(I_{3}({\bf k};m_W^2)-I_{3}({\bf k};\xi
   m_W^2)\Big) \right. \nonumber \\
   &+& \left.
   4{\bf k}^4 \Big(I_{4}({\bf k};m_W^2)- I_{4}({\bf k};\xi
   m_W^2)\Big)
    \right. \nonumber \\
   &+& \left.
    15m_W^2 I_{3}({\bf k};m_W^2)
   + 12{\bf k}^{2}m_W^2 I_{4}({\bf k};m_W^2)
   \right]
   \Big\}, \nonumber \\
\label{auto001aI}
\eea
where hereafter $\Pi^{00 \ gb} \equiv \Pi^{gb}$  for gauge boson self-energies.

Using Eqs.~(\ref{highT}) and (\ref{highT00})
into Eq.~(\ref{auto001aI}), including all contributions from gauge bosons in
the loop and keeping only the leading terms we get
\begin{eqnarray}
   \Pi^{W}_{a} &=&
 2 g^2 T^2 (\xi + 1)
- \frac{(2+\xi ^{3/2})g^2 T}{4\pi\left(g^2+g'^2\right)} \nonumber \\
&\times &\left(
\left(g^2+g'^2\right) m_W
+g^2 m_Z
+g'^2 m_\gamma
\right)
\nonumber \\
&-& \frac{g^2 (eB)^2}{128 \pi m_W^2} \left( \frac{4 T}{3 m_W}
\left(2-\frac{23 }{\xi^{1/2} }\right)
+\frac{7 \zeta(3)m_W^2}{\pi ^3 T^2} \right).   \nonumber \\
\label{auto00Wa}
\end{eqnarray}

We now turn to compute the expressions for $\Pi^{W-Z(\gamma )-S}_{b}$ depicted
in Fig.~\ref{fig2}(b). For the case where the external particle is a $W$-boson
and the internal ones are a neutral gauge boson and a charged
Goldstone boson, its explicit expression is
\begin{eqnarray}
   \Pi^{\mu \nu}_{b}(q)&=&\frac{g^2g'^2}{g^2+g'^2} \left\{
   \begin{array}{c}
   (m_Z^2-m_W^2) \\
   m_W^2
   \end{array}
   \right\}  \nonumber \\
   &\times&\int \frac{d^4 k}{(2\pi)^4} g_{\mu\rho}G^{\rho\sigma}(k)
   g_{\nu\sigma}D_B(k-q),  \label{auto1b}
\end{eqnarray}
where the upper (lower) case corresponds to an internal $Z$
($\gamma$)-line. Notice that since the net charge flowing
in the loop is not zero, the phase
factor referred to in Eq.~(\ref{phase}) does not vanish and should in
principle be also included in the computation of this self-energy.
Nevertheless, as we are interested in this expression to compute the
contribution of the ring diagrams to the EP, that in turn
is represented by closed loops, the phase factor becomes the identity and
therefore it does not need to be computed for individual self-energies. The
same applies for the rest of the self-energy diagrams where there is a net
charge flowing in the loop and consequently, we do not consider computing
their corresponding phases.

Using the Euclidean version of Eq.~(\ref{scalpropweak}) and
Eq.~(\ref{bospropweak}), we have
\bea
   \Pi^{W-Z(\gamma )-S}_{b} &&= -\frac{g^2g'^2}{g^2+g'^2} \nonumber \\
  && \hspace{-2.1cm} \times \left\{
   \begin{array}{c}
      (m_Z^2-m_W^2)\\
              m_W^2
   \end{array}
   \right\}
   T \sum_n \int \frac{d^3 k}{(2\pi)^3} \nonumber \\
&&  \hspace{-2.1cm} \times
   \Big\{
   m_{Z(\gamma)}^2 I_{11}({\bf
     k},0;m_i^2,m_{Z(\gamma)}^2) \nonumber \\
&& \hspace{-2.1cm} + \omega_n^2 \left(I_{11}({\bf
     k},0;m_i^2,m_{Z(\gamma)}^2)
   -I_{11}({\bf k},0;m_i^2,\xi m_{Z(\gamma)}^2 \right)
   \nonumber \\
&&\hspace{-2.1cm} + \frac{(eB)^2}{3}\Big[
4{\bf k}^2 m_{Z(\gamma)}^2 I_{41}({\bf
  k},0;m_i^2,m_{Z(\gamma)}^2)
 \nonumber \\
&& \hspace{-2.1cm}
  -3m_{Z(\gamma)}^2 I_{31}({\bf
    k},0;m_i^2,m_{Z(\gamma)}^2)
    \nonumber \\
&& \hspace{-2.1cm}   -3\omega_n^2 (I_{31}({\bf
      k},0;m_i^2,m_{Z(\gamma)}^2)-I_{31}({\bf
      k},0;m_i^2,\xi m_{Z(\gamma)}^2)) \nonumber \\
&& \hspace{-2.1cm}  + 4\omega_n^2\mathbf{k}^2 (I_{41}({\bf
      k},0;m_i^2,m_{Z(\gamma)}^2)
   -I_{41}({\bf k},0;m_i^2,\xi m_{Z(\gamma)}^2))
   \Big]\Big\}. \nonumber \\
\label{auto00Wb}
\eea

The integrand in Eq.~(\ref{auto00Wb}) contains terms whose general form is
\begin{eqnarray}
\label{iab}
   I_{\alpha\beta}(\mathbf{k},\mathbf{q};m^2,\xi m^2)&\equiv&\frac{1}
   {[\omega_n^2+\mathbf{k}^2+m^2]^\alpha}
   \nonumber \\
   &\times&\frac{1}{[\omega_n^2+ {(\mathbf{k-q})}^2+ \xi m^{2}]^\beta}.
   \nonumber \\
\end{eqnarray}
We make use of the Feynman parametrization to write $I_{\alpha\beta}$ as
\begin{eqnarray}
\label{feynman}
   I_{\alpha\beta}(\mathbf{k},\mathbf{q};m^2,\xi m^2)&=&\frac{
   \Gamma(\alpha+\beta)}{\Gamma(\alpha)\Gamma(\beta)}  \nonumber\\
   &\times&\int_0^1 \frac{dx \ x^{\alpha-1} (1-x)^{\beta-1}} {[\omega_n^2+
   \mathbf{k^{\prime}}^2(x)+{m^{\prime}}^2(x)]^{\alpha+\beta}},  \nonumber \\
\end{eqnarray}
where
\begin{eqnarray}
   \mathbf{k'(x)}&=& \mathbf{k}-(1-x)\mathbf{q}  \nonumber \\
   {m'^2(x)}&=& m^{2}_{W}(\xi + x - \xi x)+x(1-x)\mathbf{q}^2,
\label{defsfeynman}
\end{eqnarray}
and $\Gamma$ is the Gamma function. Notice that according to our
findings in Ref.~\cite{mishaps}, this
parametrization is allowed since we are pursuing a description in the infrared
limit. To carry out the sum over Matsubara frequencies
together with the integration in Eq.~(\ref{auto00Wb}), once again we resort
to the results in Ref.~\cite{Bedingham} (see also Ref.~\cite{sap}). The
explicit result, generalized to include the fermion case is
\bea
   T\sum_{n} \int \frac{d^d k}{(2\pi)^d}
   {\bf k}^{2a}{\omega^{2t}_{n}}
   I_{\alpha \beta}({\bf k},{\bf q};m^2,\xi m^2)&=&
   \frac{(2T) }{(4\pi)^{d/2}} \nonumber \\
   && \hspace{-5cm}
   \times\frac{(2\pi T)^{d+2a+2t-2(\alpha+\beta)}}{\Gamma(\alpha)\Gamma(\beta)}
   \frac{\Gamma(\frac{d}{2}+a)}{\Gamma(\frac{d}{2})}\mu^{2\epsilon}
   \nonumber\\
   &&\hspace{-5cm}\times
   \sum_{j=0}^\infty \frac{(-1)^j}{j!}
   \zeta(2(j+\alpha+\beta-t-\frac{d}{2}-a),Z)
   \nonumber \\
   && \hspace{-5cm} \times \Gamma(j+\alpha+\beta-\frac{d}{2}-a)
   \nonumber \\
   && \hspace{-5cm} \times
   \int_0^1 dx \ x^{\alpha-1} \ (1-x)^{\beta-1}
   \left(\frac{m'(x)}{2\pi T}\right)^{2j},
\label{highTT}
\eea
where for fermions the sum runs over all integers and $Z=1/2$. For the terms
involving the $n=0$ Matsubara frequency for bosons, we use the result
\bea
   T \int \frac{d^d k}{(2\pi)^d}{\bf k}^{2a}
   I_{\alpha\beta}({\bf k},{\bf q};m^2,\xi m^2)&=&
   \frac{T}{(4\pi)^{d/2}} \nonumber \\
&&   \hspace{-5cm} \times \frac{\Gamma(\frac{d}{2}+a)}{\Gamma(\frac{d}{2})}
   \frac{\Gamma(\alpha +\beta-\frac{d}{2})}{\Gamma(\alpha)\Gamma(\beta)}
   \nonumber \\
&&   \hspace{-5cm} \times
   \int_0^1 dx \ x^{\alpha-1} \ (1-x)^{\beta-1}
   \left(\frac{1}{m'(x)} \right)^{2\alpha+2\beta-d+2a}. \nonumber \\
\label{highTT0}
\eea

Using Eq.~(\ref{highTT}) for the terms with $n\neq 0$ and Eq.~(\ref{highTT0})
for $n=0$, and after considering the contributions from all gauge boson fields
running in the loop, we get
\begin{eqnarray}
\Pi^{W}_{b} &=&
-\frac{\left(g^2+g'^2\right)  m_W^2}{8 \pi ^2} \ln
\left(\frac{m_W}{T}\right)
-\frac{T m_W^2}{4 \pi \left(g^2+g'^2\right) } \nonumber \\
&& \hspace{-1cm} \times \left(
\frac{g^2 \left(g^2+g'^2\right) }{m_4+ m_W}
+\frac{g'^4 }{m_1+m_Z}
+\frac{g^2 g'^2}{ m_1+m_\gamma}\right) \nonumber \\
&& \hspace{-1cm} +(eB)^2 T
\left(
-\frac{g^2 m_W^2}{48 \pi  m_W (m_4+m_W)^4}\left(
\frac{ m_4^2 }{   m_W^2 }
+\frac{4  m_4 }{   m_W }
+1
\right)
\right. \nonumber \\
&& \hspace{-1cm} \left.
-\frac{7  g^2 m_4 m_W^2}{32 \pi  m_W^2
  (m_4+m_W)^3}\left(
\frac{ m_4 }{ m_W }
+3
\right)
\right. \nonumber \\
&& \hspace{-1cm} \left.
+\frac{g^2 m_W^2}{8 \pi  m_W^3}\left(
\frac{1}{ (m_4+m_W)^2}
-\frac{1}{\xi^{1/2} \left(m_4+\xi^{1/2} m_W\right)^2}
\right)
\right. \nonumber \\
&& \hspace{-1cm} \left.
-\frac{g'^4 m_W^2}{48\pi  m_1 \left(g^2+g'^2\right)
  (m_1+m_Z)^4}\left(
1
+\frac{4 m_Z}{m_1}
+\frac{m_Z^2}{m_1^2}
\right)
\right. \nonumber \\
&& \hspace{-1cm} \left.
+\frac{g'^4 m_W^2}{32 \left(g^2+g'^2\right)\pi  m_1^2
  (m_1+m_Z)^3}\left(
3
+\frac{m_Z}{m_1}
\right)
\right. \nonumber \\
&& \hspace{-1cm} \left.
-\frac{g^2 g'^2 m_W^2}{48 \left(g^2+g'^2\right)\pi  m_1
  (m_1+m_\gamma)^4}\left(
1
+\frac{4 m_\gamma}{m_1}
+\frac{m_\gamma^2}{m_1^2}
\right)
\right. \nonumber \\
&& \hspace{-1cm} \left.
+\frac{g^2 g'^2 m_W^2}{32 \left(g^2+g'^2\right) \pi  m_1^2
  (m_1+m_\gamma)^3}\left(
3
+\frac{m_\gamma}{m_1}\right)
\right)
\nonumber \\
&& \hspace{-1cm} +\frac{(eB)^2 m_W^2 \zeta(5)}{2048 \pi ^6 T^4}
\left(g'^4-15 g^4 - 14 g^2 g'^2  \right).
\label{W-Z}
\end{eqnarray}

The rest of the diagrams depicted in Fig.~\ref{fig2} can be computed in the
same manner and the result is given in Appendix B. For the expressions
corresponding to diagrams $(a)$ -- $(g)$
in Fig.~\ref{fig2}, some remarks are in order:
First, notice that terms of order $m_{gb}/T$, where $m_{gb}$
stands for any of the gauge boson masses,
are proportional to $v/T$ and thus, as in the discussion of the
scalar self-energy, these terms are small when the position of the
minimum in the broken phase is small compared to the critical temperature.
Second, when the gauge boson masses appear in the denominators throughout, we
should keep in mind that the analysis is valid in the broken phase, where the
vacuum expectation value of the Higgs field is different from zero. However,
since the analysis considers the effective potential as a function of $v\geq
0$, we will implement this restriction by substituting the gauge boson mass by
the thermal ones, when this mass appears in denominators. This is meant to
regulate the singularity at $v=0$ and does not make a
numerical difference while allowing us to treat the effective potential
as a function in the domain $v \geq 0$. The gauge boson
thermal masses are defined by
\begin{eqnarray}
   \tilde{m}_W&=&\sqrt{m_W^2+(\Pi^Q_W)_1},\nonumber\\
   \tilde{m}_Z&=&\sqrt{m_Z^2+(\Pi^Q_Z)_1},\nonumber\\
   \tilde{m}_\gamma&=&\sqrt{(\Pi^Q_\gamma)_1},
\label{massTZF}
\end{eqnarray}
where $(\Pi^Q_W)_1$, $(\Pi^Q_Z)_1$ and $(\Pi^Q_\gamma)_1$ are the leading
terms of the gauge boson self-energies made up by adding the corresponding
terms in Appendix B for each gauge boson. These leading terms turn out to be
gauge parameter independent and are explicitly given by
\begin{eqnarray}
   (\Pi^Q_W)_1&=&\frac{11}{6}g^2T^2\nonumber\\
   (\Pi^Q_Z)_1&=&\frac{11}{6}\frac{(g^4+g'^4)}{(g^2+g'^2)}T^2,\nonumber\\
   (\Pi^Q_\gamma)_1&=&\frac{11}{3}\frac{g^2g'^2}{(g^2+g'^2)}T^2.
\label{PiTZF1}
\end{eqnarray}
As is described in Ref.~\cite{nosotros}, the other non zero components of the
gauge boson self-energy are negligible
\begin{eqnarray}
   \Pi^{11}=\Pi^{22}, \ \Pi^{33}&\sim& {\mathcal{O}}(m_i^2)
   \nonumber \\
   \Pi^{03}=\Pi^{30}&\sim& {\mathcal{O}}(eB).
\label{negligible}
\end{eqnarray}
From Eqs.~(\ref{pibosonu1}) and ~(\ref{PiQM}), we see that since
$\Pi^M=-\Pi^{11}$ then $\Pi^{00}\simeq\Pi^Q$, and thus
\begin{eqnarray}
   \Pi^{\mu\nu}\simeq \Pi^{00}Q^{\mu\nu}.
\label{PiandPiQ}
\end{eqnarray}

\section{Effective Potential}\label{V}

\subsection{One-loop}

In the standard model the tree level potential is
\begin{eqnarray}
   V_{\mathrm{tree}}(v)= -\frac{1}{2} c^2 v^2 + \frac{1}{4} \lambda v^4.
\label{pothiggs}
\end{eqnarray}
To one loop, the EP receives contributions from each
sector, namely
\begin{eqnarray}
   V^{(1)}(v)=V^{(1)}_{H}(v)+V^{(1)}_{f}(v)+V^{(1)}_{\mathrm{gb}}(v)+V^{(1)}_{\mathrm{FP}}(v),
\label{oneloop}
\end{eqnarray}
where in general each one of these contributions is given by
\begin{eqnarray}
   V^{(1)}(v)= \frac{T}{2} \sum_n\int\frac{d^3k}{(2\pi)^3}
   {\mbox{Tr}}\left(\ln
   \left[D(\omega_n,\mathbf{k})^{-1}\right] \right),  \label{onelooptr}
\end{eqnarray}
with $D$ standing for either the scalar, fermion, gauge boson or ghost
propagator,
and the trace is taken over all internal indices. We first discuss the thermal
contributions coming from each sector,
as described above, and then we implement the renormalization procedure
of the temperature independent contributions.

In the weak field limit, the contribution from the Higgs sector is given by
\bea
   V^{(1)}_{H}&=&\sum_{i=1}^4 \frac{T}{2}
   \sum_n\int\frac{d^3k}{(2\pi)^3}\ln
   [D_{B}^{-1}(\omega_n,{\bf k})] \nonumber \\
   & & \hspace{-1cm} \simeq \sum_{i=1}^4 \frac{T}{2} \sum_n\int\frac{d^3k}{(2\pi)^3}
   \Big\{
   \ln(\omega_n^2+{\mathbf{k}}^2+m_i^2)
   \nonumber \\
   && \hspace{-1cm} + (e B)^2 \left[
   \frac{1}{(\omega_n^2+{\mathbf{k}}^2+m_i^2)^2}
   -\frac{2(k_\perp^2)}{(\omega_n^2+{\mathbf{k}}^2+m_i^2)^3}
   \right]\Big\}, \nonumber \\
   \label{V1ap}
\eea
where the contributions from all scalars have been accounted for.

The first term in Eq.~(\ref{V1ap}) represents the lowest
order contribution to the EP at finite temperature and zero
external magnetic field, usually referred to as the boson \textit{ideal gas}
contribution~\cite{LeBellac}. The thermal piece of this
contribution is given by~\cite{Dolan}
\bea
    V^{(1)T\neq 0}_{H}
   &\simeq&\sum_{i=1}^4\left( -\frac{\pi^2T^4}{90}
   +\frac{m_i^2T^2}{24}-\frac{m_i^3T}{12\pi} \right.\nonumber\\
   &-&\left.
   \frac{m_i^4}{32\pi^2}
   \ln\left(\frac{m_i}{4\pi T}\right)+
   {\mathcal{O}}(m_i^4)\right).
   \label{idealgas}
\eea
Notice that there are potentially dangerous terms in Eq.~(\ref{idealgas}) that
can become imaginary for negative values of $m_i$. However as we
will show, these terms cancel when including the Higgs contribution from the
vacuum renormalization as well as the ring diagrams.

The second, $B$-dependent term in Eq.~(\ref{V1ap}) vanishes
identically~\cite{nosotros}. Therefore, to one-loop order, the
thermal contribution to the EP in the weak field case from the Higgs sector
is independent of $e B$ and is given by Eq.~(\ref{idealgas}).

In the weak field limit, the contribution from the fermion sector is given
by
\bea
   V^{(1)}_{f}&=&N_c
   T  \sum_n\int\frac{d^3k}{(2\pi)^3}\ln
   [S_{B}^{-1}(\omega_n,{\bf k};{m_t})] \nonumber\\
    &\simeq& N_c
   2T\sum_n \int\frac{d^3k}{(2\pi)^3}
   \Big\{ \ln[\omega_n^2+{\bf k}^2+{m_t}^2] \nonumber \\
   &+&
   2(e B)^2 \frac{\omega_n^2+k_3^2+{m_t}^2}
                       {(\omega_n^2+{\bf k}^2+{m_t}^2)^3}
   \Big\},
\label{oneloopfer}
\eea
where $N_c=3$ is the number of colors. We emphasize that the
only fermion mass we keep in the analysis is the top mass $m_t$.

The first term in Eq.~(\ref{oneloopfer}) represents the fermion \textit{ideal
gas} contribution~\cite{LeBellac}, whose thermal part
is explicitly given by
\bea
   V^{(1)T\neq 0}_{f}
   &\simeq & 3\left[-7\frac{\pi^2 T^4}{180}
   +\frac{{m_t}^2T^2}{12} \right. \nonumber \\
   &+& \left. \frac{{m_t}^4}{16 \pi^2}
   \ln\left(\frac{{m_t}^2}{T^2}\right)
   +{\mathcal{O}}({m_t}^4)\right].
\label{idealgasfer}
\eea

The second term in Eq.~(\ref{oneloopfer}) is subdominant, after taking care
of renormalization as shown in Ref.~\cite{sap}.
Therefore, to one-loop order, the thermal contribution to the
EP in the weak field limit from the fermion sector is only
given by Eq.~(\ref{idealgasfer}).

In the gauge boson and ghost sectors, first
note that in the broken phase the charged fields do
couple to the external magnetic field. Therefore, their
contribution to the EP at one-loop contains both magnetic field
independent and dependent terms. For the former, the ghost field
contributions cancel the spurious degrees of freedom arising in the
gauge sector when working in an arbitrary covariant gauge. In the latter,
the charged ghost fields contribute as regular scalar
fields with the opposite sign and vanishes identically~\cite{nosotros} [see discussion after Eq.~(\ref{idealgas})]. Altogether, the magnetic field
dependent contribution from these sectors comes only from the
magnetic field dependent pieces of the W boson propagator.

Therefore, the thermal part of the contribution to the EP
from the gauge boson sector is given explicitly by
\begin{widetext}
\bea
   V^{(1)T\neq 0}_{\mathrm{gb}} + V^{(1)T\neq 0}_{\mathrm{FP}}&=&
   -11 \frac{\pi^2T^4}{90}
   +3\frac{(2m_W^2+m_Z^2)T^2}{24}
   -3\frac{(2m_W^3+m_Z^3)T}{12\pi}
   -6\frac{m_W^4}{32\pi^2}\ln
   \left(\frac{m_W}{4\pi T}\right)
   -3\frac{m_Z^4}{32\pi^2}\ln\left(\frac{m_Z}{4\pi T}\right) \nonumber\\
   &&\hspace{-1cm} -\frac{(eB)^2}{256 \pi^2} \left[
   {\mathcal P}_0(\xi)
   +{\mathcal P}_1(\xi)\ln\left(\frac{\tilde{m}_W}{T}\right)
   +\pi {\mathcal P}_2(\xi) \frac{T}{\tilde{m}_W}
   +\frac{\zeta(3)}{\pi^2}{\mathcal P}_3(\xi) \frac{m_W^2}{T^2}
   +\frac{\zeta(5)}{\pi^4} \frac{m_W^4}{T^4}
   \right],
\label{idealgasgbT}
\eea
\end{widetext}
where we have maintained the leading magnetic field dependent term
and we defined
\bea
\label{polinomios}
{\mathcal P}_0(\xi) &=& \frac{32(1 - 5\xi)}{3\xi}, \nonumber \\
{\mathcal P}_1(\xi) &=& \frac{64(3+7\xi)}{\xi}, \nonumber \\
{\mathcal P}_2(\xi) &=& \frac{8}{3\xi(1+\xi^{1/2})} \left(14+53\xi^{1/2}
+361\xi \right. \nonumber \\
&&\left.
+170\xi^{3/2}-120\xi^2+\xi^{5/2}+\xi^3\right), \nonumber \\
{\mathcal P}_3(\xi) &=& \frac{8(1+16\xi)}{3\xi}.
\eea
Note that the factors in front of each magnetic field independent contribution
in Eq.~(\ref{idealgasgbT}) correspond to
the two $W^{\prime}$s, the $Z$ and the photon
polarizations~\cite{Carrington}. Also, notice that in the terms
that come within the contributions proportional to the magnetic field squared,
we have replaced the $W$-boson mass by its thermal one within the argument of
the logarithmic function and in the term where that mass appears in the
denominator. This is because, as previously discussed, although
the analysis is strictly speaking only valid in the broken phase, since we
consider the effective potential as a function of $v\geq 0$, to avoid
divergences near $v\simeq 0$ we replace the gauge boson mass by the thermal
one.

To implement the renormalization procedure, notice that the $T=0$ one-loop
EP from Eq.~(\ref{onelooptr}) involves the integral
\bea
   {\mathcal{V}}(m)=\frac{1}{2}\int\frac{d^3k}{(2\pi)^3}\sqrt{k^2+m^2},
   \label{T=0}
\eea
where $m$ stands for the mass of any one of the contributing species. This
integral diverges and needs to be regularized. We do so by means of
introducing the ultraviolet cutoff $\Lambda$ and thus
\bea
   {\mathcal{V}}(m;\Lambda)&=&\frac{1}{4\pi^2}\int_0^\Lambda
   k^2\sqrt{k^2+m^2}dk\nonumber\\
   &\stackrel{\Lambda\rightarrow\infty}{\rightarrow}&
   \frac{1}{16\pi^2}\left[\Lambda^4 + m^2\Lambda^2 + \frac{m^4}{8}
   \right.\nonumber\\
   &+&\left.\frac{m^4}{2}\ln\left(\frac{m}{2\Lambda}\right)\right].
   \label{T=0,L}
\eea
Therefore, the $T=0$ part of the one-loop potential can be written as
\bea
   V^{(1)T=0}&=&{\mathcal{V}}(m_H;\Lambda) +
   \sum_{i=1}^3{\mathcal{V}}(m_i;\Lambda)
   + 8{\mathcal{V}}(m_W;\Lambda)\nonumber\\
   &+& 4{\mathcal{V}}(m_Z;\Lambda)
   - 12{\mathcal{V}}(m_t;\Lambda)
   -\sum_{i=1}^3{\mathcal{V}}(m_{\eta_i};\Lambda),\nonumber \\
   \label{VeffT=0}
\eea
where the factors in front of each term account for the degrees of freedom
of the corresponding species and the masses depend on $v$. Notice that since
we are working in arbitrary covariant gauge the unphysical degrees of freedom
for gauge bosons are canceled by the contributions of the ghost fields, which
leaves Eq.~(\ref{VeffT=0}) with only the physical degrees of freedom. We
emphasize that the only fermion we consider is the top quark. Since the theory
is renormalizable, it should be possible to absorb the $\Lambda$-dependent
terms with the introduction of suitable counterterms that maintain the form of
the three-level potential. Therefore, the general structure of the $T=0$
effective potential up to one-loop order, after renormalization, should be
\bea
   V^{(1)T=0}_{\mathrm{ren}}(v)&=&-\frac{1}{2}c^2v^2 + \frac{1}{4}\lambda v^4\nonumber\\
   &+&\frac{a(\Lambda)-\delta c^2}{2}v^2 +
   \frac{b(\Lambda)+\delta\lambda}{4}v^4\nonumber\\
   &+& V^{(1)T=0},
   \label{VeffT=0ren}
\eea
where the coefficients $a(\Lambda)$ and $b(\Lambda)$ are introduced to cancel
the $\Lambda$-dependent terms in $V^{(1)T=0}$ and the coefficients
$\delta c^2$ and $\delta\lambda$ take care of possible finite corrections
of the $v^2$ and $v^4$ terms, respectively.
As the renormalization condition, we require that the minimum of
$V^{(1)T=0}_{ren}(v)$ remains at its classical value, namely
\bea
   \left.\frac{dV^{(1)T=0}_{\mathrm{ren}}}{dv}\right|_{v=v_0}=0.
   \label{condition}
\eea
\begin{figure}[t!] 
\vspace{0.5cm}
{\centering
{\epsfig{file=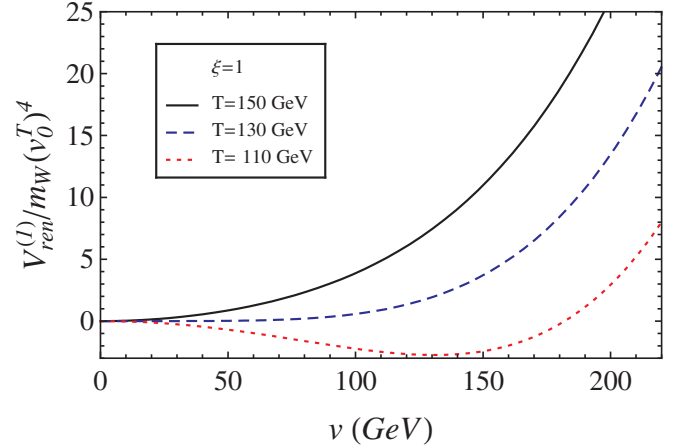, width=1\columnwidth}}
\par}
\caption{Renormalized one-loop effective potential for $\xi=1$ and different
values of $T$, subtracting $v$ independent terms and dividing by the
fourth power of the $v$-dependent $W$ mass evaluated at $v=v_0^T$, the value of
the vacuum expectation value that minimizes de effective potential after the
phase transition is completed for a temperature $T=110$ GeV.
In this gauge the phase transition occurs at $T_c=128.9$ GeV and is
second order.}
\label{fig3}
\end{figure}
Introducing suitable constant terms to make the arguments of the logarithms be
dimensionless and disregarding additive constants and terms explicitly
proportional to $\lambda^2$, we get
\bea
   V^{(1)T=0}_{\mathrm{ren}}(v)&=&-\left(1+\frac{3\lambda}{16\pi^2}+
   \frac{(g^2+g'^2)\xi}{128\pi^2} + 2\frac{g^2\xi}{128\pi^2}\right)
   \nonumber\\
   &\times&
   \frac{c^2v^2}{2}\nonumber\\
   &+&\left(\lambda + \frac{3f^4}{32\pi^2} - \frac{3g^4}{256\pi^2}
   -\frac{3(g^2+g'^2)^2}{512\pi^2}\right. \nonumber\\
   &-&\left.
    \frac{g^4\xi^2}{256\pi^2}
   - \frac{(g^2+g'^2)^2\xi^2}{512\pi^2}
   \right)\frac{v^4}{4}\nonumber\\
   &+&\frac{m_H^4}{64\pi^2}\ln\left(\frac{m_H^2}{4c^2}\right)
   \nonumber\\
   &+&\sum_{i=1}^3\frac{m_i^4}{64\pi^2}\ln
   \left(\frac{m_i^2}{4c^2}\right)
   \nonumber\\
   &+&6\frac{m_W^4}{64\pi^2}\ln\left(\frac{m_W^2}{m_W(v_0)^2}\right)
   \nonumber\\
   &+&3\frac{m_Z^4}{64\pi^2}\ln\left(\frac{m_Z^2}{m_Z(v_0)^2}\right)
   \nonumber\\
   &-&12\frac{m_f^4}{64\pi^2}\ln\left(\frac{m_f^2}{m_f(v_0)^2}\right).
\eea
Therefore, after renormalization, the 1-loop EP at finite
temperature is given by
\bea
   V^{(1)}_{\mathrm{ren}}(v)&=&-\left(1+\frac{3\lambda}{16\pi^2}+
   \frac{(g^2+g'^2)\xi}{128\pi^2} + 2\frac{g^2\xi}{128\pi^2}\right)
   \frac{c^2v^2}{2}\nonumber\\
   &+&\left(\lambda + \frac{3f^4}{32\pi^2} - \frac{3g^4}{256\pi^2}\right.
   \nonumber\\
   &-&\left.
   \frac{3(g^2+g'^2)^2}{512\pi^2} - \frac{g^4\xi^2}{256\pi^2}
   - \frac{(g^2+g'^2)^2\xi^2}{512\pi^2}
   \right)\frac{v^4}{4}\nonumber\\
   &+&\sum_{i=1}^4\left[\frac{m_i^2T^2}{24} - \frac{m_i^3T}{12\pi}
   +\frac{m_i^4}{64\pi^2}\ln\left(\frac{(4\pi T)^2}{c^2}\right)\right]
   \nonumber\\
   &+&3\frac{m_t^2T^2}{12} -3\frac{m_t^4}{16\pi^2}
   \ln\left(\frac{T^2}{m_t^2(v_0)}\right)\nonumber\\
   &+&3\frac{(2m_W^2+m_Z^2)T^2}{24} - 3\frac{(2m_W^3+m_Z^3)T}{12\pi}
   \nonumber\\
   &+&6\frac{m_W^4}{64\pi^2}\ln\left(\frac{(4\pi T)^2}{m_W(v_0)^2}\right)
   \nonumber\\
   &+&3\frac{m_Z^4}{64\pi^2}\ln\left(\frac{(4\pi T)^2}{m_Z(v_0)^2}\right)
   \nonumber\\
   &-&\frac{(eB)^2}{256 \pi}{\mathcal P}_2(\xi) \frac{T}{\tilde{m}_W},
   \label{1loopVfinT}
\eea
where we have kept only the leading magnetic filed dependent term, with
${\mathcal P}_2$ as defined in Eq.~(\ref{polinomios}) and the masses without
an argument are the $v$-dependent masses. As pointed out, the dangerous
dependence of the scalar masses in the arguments of the logarithmic
functions have disappeared. However, to cancel the cubic terms in the scalar
masses one needs to go to the next order, namely, to the ring diagrams.
Let us for the time being ignore the terms proportional to cubic scalar
masses and explore the properties of the renormalized effective potential at
1-loop. Figure~\ref{fig3} shows $V^{(1)}_{\mathrm{ren}}(v)$ for $\xi=1$ and different
values of $T$, subtracting $v$ independent terms and dividing by the
fourth power of the $v$-dependent $W$ mass evaluated at $v=v_0^T$, the value of
the vacuum expectation value that minimizes de effective potential after the
phase transition is completed for a temperature $T=110$ GeV. We can see that
the phase transition is second order and in this
gauge it happens for $T_c=128.9$ GeV. To illustrate the gauge dependence of the
phase transition parameters, Fig.~\ref{fig4} shows $V^{(1)}_{\mathrm{ren}}(v)$ for a low
temperature $T=110$ GeV, for which the phase transition has been completed,
for different values of the gauge parameter $\xi$. Notice the
dependence on the gauge parameter of the broken phase minimum
at this constant temperature, whose numerical values are shown on the second
column of table~I. The dependence on the gauge parameter
of the critical temperature is milder, as shown on the first column of table~I.
\begin{figure}[t!] 
\vspace{0.5cm}
{\centering
{\epsfig{file=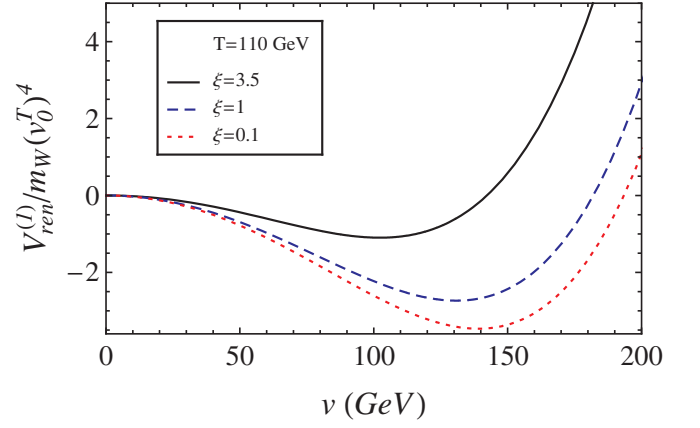, width=1\columnwidth}}
\par}
\caption{The renormalized one-loop effective potential normalized
by the fourth power of the $v$ dependent $W$-mass evaluated at $v=v_0^T$, the
value of the vacuum expectation value that minimizes de effective potential
after the phase transition is completed for a temperature $T=110$ GeV and
$\xi=1$, as a function  of the vacuum expectation value $v$ and for
different values of the gauge parameter $\xi$. At this temperature, the
position of the minimum $v_0$ shows an somewhat significant dependence on the
gauge parameter for values $\xi$ around $1$.}
\label{fig4}
\end{figure}
\begin{table}[t!]
\centering
\begin{tabular}{ccc}
\hline
   $\xi$ & $T_c$ (GeV) & $\mathrm{v_0}$ (GeV) \\ [0.5ex]
\hline
   3.5 &    120.8  &  102.3   \\
    1  &    128.9  &  130.9   \\
   0.1 &    132.4  &  139.0   \\ [1ex]
\hline
\end{tabular}
\label{tabla}
\caption{Critical temperature $T_c$ (second column) and position of the minimum
  $\mathrm{v_0}$ for the constant temperature $T=110$ GeV (third column) of the
  renormalized one-loop effective potential, after the phase transition has
  been completed. The chosen values of $\xi$ are close to $1$, Notice that the
  critical temperature is only mildly dependent on $\xi$, however the
  position of the minimum for a temperature where the phase transition has
  been completed shows a larger $\xi$ dependence.}
\end{table}
\subsection{Ring diagrams}

\begin{figure}[t!] 
{\centering
{\epsfig{file=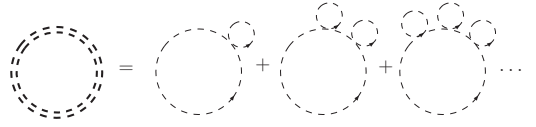, width=0.8\columnwidth}}
\par}
\caption{Schematic representation of the resummation of the ring diagrams.}
\label{fig5}
\end{figure}

It is well known that the next order correction to the EP comes from the so
called \textit{ring diagrams}. These are schematically depicted in
Fig.~\ref{fig5}. Their renormalized contribution to the EP is more clearly
found by explicitly separating the two-loop contribution~\cite{LeBellac}
coming from the scalar sector and can be written as
\bea
   V^{(\mathrm{ring})}_{H}(v)&\simeq&
   \frac{T}{2}
   \sum_n\int\frac{d^3k}{(2\pi)^3}
   \left\{(\ln[1+\Pi^{H}D^{H}]\right.\nonumber\\
   &-& \Pi^{H}D^{H})\nonumber \\
   &+& (\ln[1+\Pi^{G^0}D^{G^0}] - \Pi^{G^0}D^{G^0})\nonumber \\
   &+& \left. 2(\ln[1+\Pi^{G^c}D^{G^c}] - \Pi^{G^c}D^{G^c})
   \right\}\nonumber\\
   &+& V^{(2)}_S,
\label{ringhiggs}
\eea
where $\Pi^{H}$, $\Pi^{G^0}$ and $\Pi^{G^c}$ are the Higgs, neutral and
charged scalar self-energies in the presence of the magnetic field,
$D^{H}$, $D^{G^0}$ and $D^{G^c}$ their corresponding propagators and
$V^{(2)}_S$ the contribution to the two-loop effective potential
coming exclusively from the scalar sector. The factor 2 accounts for the
two charged scalar degrees of freedom. As written, Eq.~(\ref{ringhiggs})
deserves some comments: First, the dominant contribution comes from the
Matsubara frequency with $n=0$. Second, ultraviolet divergences are canceled
explicity. Third, the full two-loop contribution involving the scalar
sector contains diagrams with particles other than scalars. For the purposes
of the present analysis, we consider only the subclass of diagrams that
contain only scalars and thus we get
\bea
   V^{(2)}_S&\simeq&\frac{\lambda}{24}T^4 - \frac{\lambda}{16\pi}
   \left(m_H + m_3 + 2m_1\right)T^3\nonumber\\
   &-&(eB)^2\frac{\lambda}{192\pi m_1^3}T^3,
\label{V2scalaraprox}
\eea
where we keep only the leading contributions, within the hierarchy of energy
scales considered. The potentially dangerous terms with odd powers of
the scalar masses in Eq.~(\ref{V2scalaraprox}) exactly cancel
similar terms coming form the integral in Eq.~(\ref{ringhiggs}), under the
approximation that the self-energies involve only the scalar contribution. The
full proof of this cancellation in the absence of magnetic fields has been
treated in detail for the standard model in Ref.~\cite{arnold}. In the
presence of an external magnetic field, it has been shown in the linear sigma
model in Ref.~\cite{sigmapaper}. Within our approximation, this result shows
that considering only scalar contributions, this cancellation also happens in
the standard model with a magnetic field. A full proof is currently under
way~\cite{jorge}.

The dominant contribution in Eq.~(\ref{ringhiggs}) comes from the mode
$n=0$. The explicit expression for Eq.~(\ref{ringhiggs}) is~\cite{nosotros}
\bea
   V^{(\mathrm{ring})}_{H} &=&
   \frac{\lambda}{24}T^4
   - \frac{T}{12\pi}\left\{\left(m_H^2+\Pi^{H} \right)^{3/2} - m_H^3
   \right.\nonumber\\
   &+&\left(m_3^2+\Pi^{G^0} \right)^{3/2} - m_3^3
   +2\left[\left(m_1^2+\Pi^{G^c} \right)^{3/2}\right.\nonumber\\
   &-&\left.\left. m_1^3
   +\frac{(eB)^2\Pi_1}{16\left(m_1^2+
   \Pi^{G^c} \right)^{3/2}}\right]\right\}.
\label{VringBno0}
\eea

As anticipated, the cubic mass terms in Eq.~(\ref{VringBno0}) will exactly
cancel the similar terms appearing in Eq.~(\ref{1loopVfinT}) after adding both
equations.

Next, we turn to the ring contribution from the gauge boson and Faddeev-Popov
ghost sectors. The computation is simplified by recalling that the ghost
degrees of freedom cancel the leading temperature contributions from spurious
degrees of freedom arising form the covariant treatment of the gauge
fields. The contribution that depends on the
magnetic field, being subleading, is not canceled, as was the case for the
one-loop contribution. Thus we write
\begin{eqnarray}
   V^{(\mathrm{ring})}_{\mathrm{gb}}(v)&+&V^{(\mathrm{ring})}_{\mathrm{FP}}=-\frac{T}{2}\sum_{\mathrm{gb}}
   \sum_n\int\frac{d^3k}{(2\pi)^3}\nonumber\\
   &\times& \mbox{Tr} \left\{\sum_{N=2}^{\infty} \frac{1}{N}\left[%
   -\Pi_{\mu\lambda\, \mathrm{gb}}(0) G^{\lambda\nu}_{\mathrm{gb}}(\omega_n,\mathbf{k}) \right]^N
   \right\}\nonumber\\
   &+& V^{(ring)}_{charged\ FP}
\label{ringgb}
\end{eqnarray}
where the sum over $\mathrm{gb}$ runs over the four gauge boson physical
degrees of freedom and $V^{(\mathrm{ring})}_{\mathrm{charged}\ \mathrm{FP}}$ represents the subleading,
magnetic field dependent term, arising from the ring contribution of the
charged ghost fields. Notice
that this time, in contrast to the scalar case, we do not add explicitly the
two-loop contribution and consequently the sum over the index $N$ starts from
$N=2$.

We first compute the contribution from the neutral gauge bosons to
Eq.~(\ref{ringgb}). In the approximation where the particle has a small
momentum (the infrared limit), the Euclidean version of a neutral, gauge boson
propagator, for an arbitrary value of the gauge parameter $\xi$
can be written as
\begin{eqnarray}
   G_{\mu\nu}&=&  \frac{1}{(k^2+m_{\mathrm{gb}}^2)}\left\{P^L_{\mu\nu}+P^T_{\mu\nu}+\xi
   \frac{k_\mu k\nu}{(k^2 +\xi m_{\mathrm{gb}}^2)}\right\},\nonumber\\
\label{propbas}
\end{eqnarray}
where
\begin{eqnarray}
   P^T_{00}=P^T_{0i}=0& &P^T_{ij}=\delta_{ij}-\mathbf{\hat{k}}_i
   \mathbf{\hat{k}}_j \nonumber \\
   P^L_{\mu\nu}=\delta^{\mu\nu} \!\!&-&\!\! \frac{k_\mu k_\nu}{k^2}
   -P^T_{\mu\nu}.
\label{translong}
\end{eqnarray}
By using Eqs.~(\ref{PiandPiQ}) and~(\ref{translong}), it is easy
to see that the product
$\Pi_{\mu\lambda}(0)G^{\lambda\nu}(\omega_n,\mathbf{k})$ becomes
\begin{eqnarray}
   \Pi_{\mu_\lambda} G^{\lambda\nu}&=& \frac{(\Pi_{\mathrm{gb}}^{Q})_1}{(k^2+m_{\mathrm{gb}}^2)}
   \left[1+\xi\frac{(k\cdot u)^2}{(k^2 +\xi m_{\mathrm{gb}}^2)}\right]Q_\mu^\nu,
   \nonumber\\
\label{propbas2-1}
\end{eqnarray}
from where, considering the $n=0$ term and taking the trace,
we can write Eq.~(\ref{ringgb}) as
\bea
    V^{(\mathrm{ring})}_{\mathrm{neutral}\ \mathrm{\mathrm{gb}}}(v) &\simeq&\sum_{\mathrm{gb}} \frac{1}{2} T
    \int\frac{d^3k}{(2\pi)^3}
    \left[ \ln \left(1+\frac{(\Pi^{Q}_{\mathrm{gb}})_1}{{\bf k}^2+m_{\mathrm{gb}}^2}\right)
    \right.\nonumber\\
   &-&\left.\frac{(\Pi^{Q}_{\mathrm{gb}})_1}{{\bf k}^2+m_{\mathrm{gb}}^2}\right] \nonumber \\
    &=&-\frac{T}{12\pi}\left\{
    \tilde{m}_Z^3 -m_Z^3+\tilde{m}_{\gamma}^3\right\}.
\label{ringgb2}
\eea
Notice that the plasma screening effects in Eq.~(\ref{ringgb2}) are
naturally accounted for by the fact that the
thermal modification of the gauge boson mass $\tilde{m}_{\mathrm{gb}}$ ($\tilde{m}_Z$
and $\tilde{m}_\gamma$) appear. Also, the nice cancellations that took place
in the scalar sector between odd powers of the masses will not happen in this
case when adding these terms to Eq.~(\ref{1loopVfinT}). This is because in the
infrared limit there is no contribution from the gauge boson
transverse degrees of freedom, thus there is no match of the coefficients to
produce the cancellation. This is a feature of the infrared limit we are
considering. However, since the square of the gauge boson masses
are never negative, these terms do not pose a problem.

Next, we turn to the contribution of the charged gauge fields. In
the weak field limit, the product
$\Pi_{\mu\lambda}(0)G^{\lambda\nu}(\omega_n,\mathbf{k})$ can be written as
\begin{eqnarray}
   \Pi_{\mu\lambda} G^{\lambda\nu}&=& \frac{(\Pi^Q_W)_1}{k^2+m_W^2}
   \left[1+(eB)^2\left(\frac{1}{(k^2 +m^2)^2} \right. \right. \nonumber\\
   &&\left. \left. -\frac{2k^2_{\perp}}{(k^2 +m^2)^3}\right)\right]Q_\mu^\nu,
\label{propbas2}
\end{eqnarray}
Using Eq.~(\ref{propbas2}), and carrying out an expansion of the
argument of the logarithm, we can explicitly write
\bea
   \ln[1+\Pi_{\mu\lambda}G^{\lambda\nu}]
   &=& \ln\left\{1+\left[\frac{(\Pi^Q_W)_1}{{\bf
         k}^2+m_W^2}\right]Q_\mu^\nu\right\}
   \nonumber\\
   &+&\ln\left\{1+\left[\frac{(\Pi^Q_W)_1}{{\bf k}^2+m_W^2
   +(\Pi^Q_W)_1}\right.\right.
   \nonumber\\
   &\times&\left(\frac{1}{({\bf k}^2+m_W^2)^2}
   -\frac{2k^2_{\perp}}{({\bf k}^2+m_W^3)^3}\right)\nonumber\\
   &\times&\left.\left.(eB)^2\right]Q_\mu^\nu
   \right\}.
\label{ringgb3}
\eea
Using this result into Eq.~(\ref{ringgb}), considering
the $n=0$ term and taking the trace in Eq.~(\ref{ringgb}) we get
\begin{widetext}
\bea
   V^{(\mathrm{ring})}_{\mathrm{charged}\ \mathrm{gb}}(v)
   &=&
   \sum_{\mathrm{gb}}
   \frac{1}{2} T \int\frac{d^3k}{(2\pi)^3}
   \left\{\left[
   \ln\left(1+\frac{(\Pi^Q_W)_1}{{\mathbf k}^2+m_{\mathrm{gb}}^2}\right)
   -\frac{(\Pi^Q_W)_1}{{\bf k}^2+m_{\mathrm{gb}}^2}
   \right]
   \right.\nonumber\\
   &+&
   \left.
   \ln\left[1+\frac{(\Pi^Q_W)_1}{({\mathbf{k}}^2+m_{gb}^2+(\Pi^Q_W)_1)}
   \left(
   \frac{1}{({\mathbf{k}}^2+m_{\mathrm{gb}}^2+(\Pi^Q_W)_1)^2}
   -\frac{2k_\perp^2}{({\mathbf{k}}^2+m_{\mathrm{gb}}^2+(\Pi^Q_W)_1)^3}
   \right)
   (e B)^2
   \right]\right\}
   \nonumber \\
   &=&-2\frac{T}{12\pi}\left(
   \tilde{m}_W^3-m_W^3\right)
   - \frac{(eB)^2}{4\pi}\left(\frac{(\Pi^Q_W)_1}{48}\right)
   \left(\frac{T}{\tilde{m}_W^3}\right).
\label{ringgb4}
\eea
\end{widetext}
In this approximation, the plasma screening
effects in Eq.~(\ref{ringgb4}) emerge naturally as a thermal
modification of the gauge boson mass, $\tilde{m}_W$ defined in
Eqs.~(\ref{massTZF}) and~(\ref{PiTZF1}).

\begin{figure}[t!] 
\vspace{0.5cm}
{\centering
{\epsfig{file=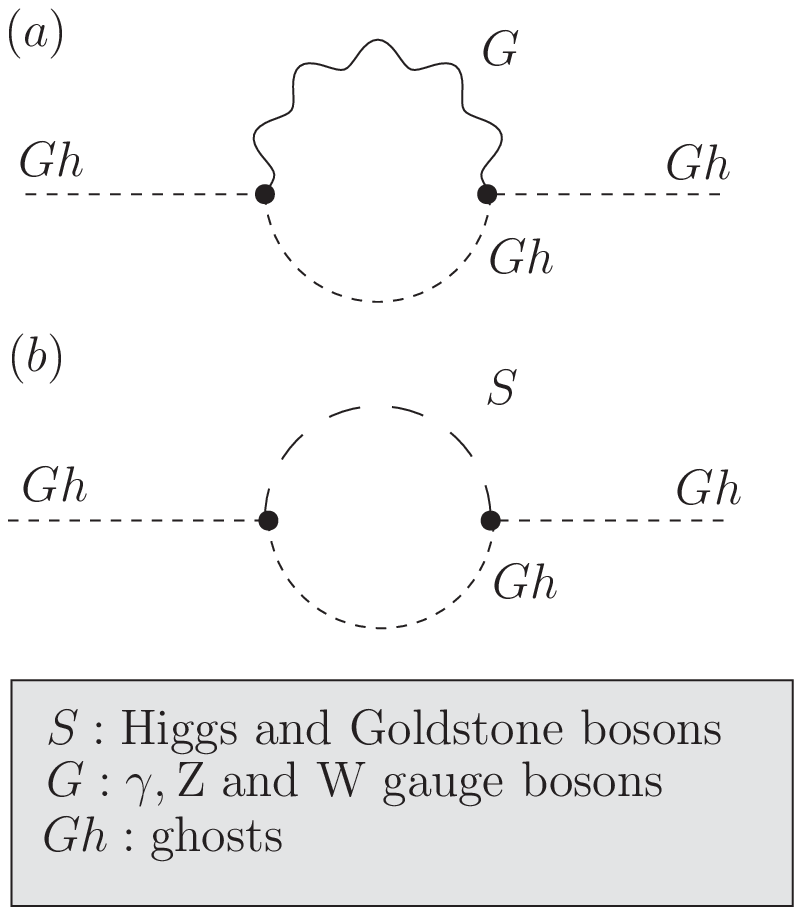, width=0.9\columnwidth}
}
\par}
\caption{Feynman diagrams for the Faddeev-Popov ghosts self-energies.
Notice that since we work in the infrared limit, the diagrams
of type \textit{(a)}, do not contribute to the ghosts self-energies.}
\label{fig6}
\end{figure}

Finally, the magnetic field dependent contribution from the charged ghost
fields to the ring potential is easily computed recalling that this
contribution is equivalent to the one coming from two scalars (one for each
$W$ field) but with opposite sign. Thus we have
\begin{eqnarray}
   V^{(\mathrm{ring})}_{\mathrm{charged}\ \mathrm{FP}}=2
   \frac{(eB)^2}{2\pi}\left(T\frac{(\Pi_{\eta_W})_1}{48(\tilde{m}_{\eta_W})^3}
   \right),
\label{ringchargedghosts}
\end{eqnarray}
where the diagram depicting the Faddeev-Popov ghosts is shown in
Fig.~\ref{fig6} and, as shown in Appendix C, the leading contribution to the
$W$-ghost self energy is given by
\begin{eqnarray}
   (\Pi_{\eta_W})_1&=&\frac{\xi^2 T m_W \mathrm{g}^2}{16\pi}\times \nonumber\\
   && \left\{ \frac{m_W}{\tilde{m}_3 + \xi^{1/2}m_W}
   -\frac{m_W}{\tilde{m}_4 + \xi^{1/2}m_W} \right. \nonumber\\
   &+&\left.\frac{\mathrm{g}^2-\mathrm{g}'^2}{\mathrm{g}(\mathrm{g}^2+\mathrm{g}'^2)^{1/2}}
   \left(\frac{m_Z}{\tilde{m}_1 + \sqrt{\xi}m_Z}\right)\right\},
\label{Pighost1}
\end{eqnarray}
and the thermal $W$($Z$)-ghost mass is given by
\begin{eqnarray}
   \tilde{m}_{\eta_{W(Z)}}&=&\sqrt{\xi m_{W(Z)}^2 + (\Pi_{\eta_{W(Z)}})_1},
\label{thermalmghost}
\end{eqnarray}
$m_W$ and $m_Z$ are the $v$-dependent $W$ and $Z$ masses, respectively and
$\tilde{m}_{i}$ ($i=1\ldots 4$) are defined in Eqs.~(\ref{higgsmass})
and~(\ref{scalarthermalmass}). Notice that in Eq.~(\ref{Pighost1}), we have
replaced $m_i$ by $\tilde{m}_i$ since, although the analysis is valid near the
broken phase minimum, we consider the effective potential as a function of
$v\geq 0$, and for small values of $\mathrm{v}$, the square of the scalar masses can
become negative.

\subsection{Effective potential up to ring order}

The final expression for the effective potential
\begin{eqnarray}
   V_{\mathrm{eff}}(v)&=&V_{\mathrm{tree}}(v)+V^{(1)}_{H}+V_f^{(1)}+V^{(1)}_{\mathrm{gb}}+V^{(1)}_{\mathrm{FP}}
   \nonumber \\
   &+& V^{(\mathrm{ring})}_{H}+V^{(\mathrm{ring})}_{\mathrm{gb}}+V^{(\mathrm{ring})}_{FP}
\label{finalpot}
\end{eqnarray}
is obtained by adding up
the results in Eqs.~(\ref{pothiggs}), (\ref{idealgas}), (\ref{idealgasfer}),
(\ref{idealgasgbT}), (\ref{VringBno0}), (\ref{ringgb4})
and~(\ref{ringchargedghosts}).

In order for the terms involving the square of the scalar bosons' thermal mass
to be real, the temperature must be such that
\begin{eqnarray}
   T>T_1&\equiv&\sqrt{\frac{16 c^2}{3\mathrm{g}^2+\mathrm{g}'^2+8\lambda+4f^2}},
\label{bound1}
\end{eqnarray}
which defines a lower bound for the temperature. A more restrictive bound is
obtained by requiring that $eB < \tilde{m}_H^2$
for the weak field expansion to work. This condition translates into the
bound
\begin{eqnarray}
   T>T_2&\equiv&\sqrt{\frac{eB+16 c^2}
   {3\mathrm{g}^2+\mathrm{g}'^2+8\lambda+4f^2}}.
\label{bound2}
\end{eqnarray}
The relevant factor that enhances the order of transition, present both in
$V^{(\mathrm{ring})}_{H}$ and $V^{(\mathrm{ring})}_{\mathrm{gb}}$, is $(eB)^2/\tilde{m}_i^3$ which
can be traced back to the boson self-energy diagrams involving a tadpole of
charged scalars in the presence of the external field.

\section{Symmetry restoration}

\label{VI}

In order to quantitatively check the effect of the magnetic field during the
EWPT, we proceed to plot $V_{\mathrm{eff}}$ as a function of the vacuum expectation
value $v$. For the analysis we use $\mathrm{g}^{\prime}=0.344$ and $\mathrm{g}=0.637$, $m_Z=91$
GeV, $m_W=80$ GeV, $f=1$, $\lambda=0.11$ which corresponds to the current
bound on the Higgs mass.

\begin{figure}[t!] 
{\centering
{\epsfig{file=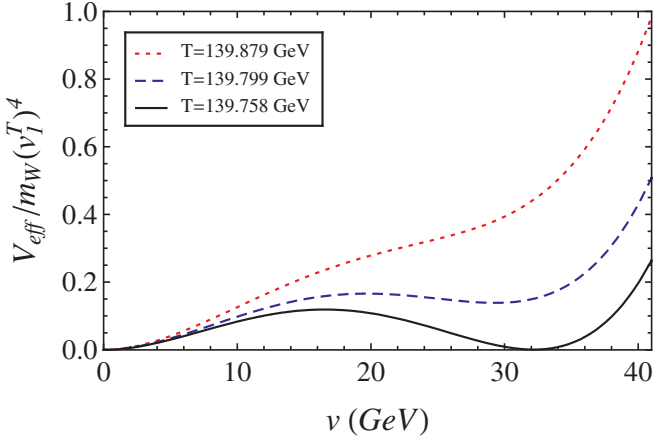, width=1\columnwidth}
}
\par}
\caption{Effective potential divided by the fourth power of the $v$-dependent
  $W$ mass evaluated at $v_1^T$ (the value where the broken phase minimum
  appears at the critical temperature in the absence of magnetic field). The
  gauge parameter $\xi=0.1$ and the value of the magnetic field is set to
  zero. The phase transition is weakly first order. Note that the
  results are consistent with the ones obtained by~\cite{sap} which are
  computed using symmetric phase degrees of freedom}.
\label{fig7}
\end{figure}

Figure~\ref{fig7} shows the effective potential in the absence of magnetic
field divided by $[m_W(v_1^T)]^4=(10.5\ {\mbox{GeV}})^4$, where $v_1^T$ is the
value where the broken phase minimum appears at the critical temperature,
which in this case happens to be $T=139.758$ GeV. The value of the gauge
parameter is $\xi=0.1$. The phase transition is weakly first order. Note that
the results are consistent with the ones obtained in Ref.~\cite{sap} which are
computed using symmetric phase degrees of freedom.

\begin{figure}[b!] 
\vspace{0.1cm}
{\centering
{\epsfig{file=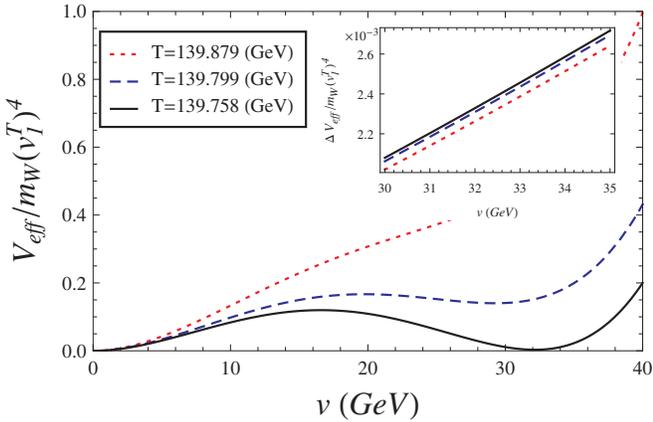, width=1\columnwidth}
}
\par}
\caption{Effective potential divided by the fourth power of the $v$-dependent
  $W$ mass evaluated at $v_1^T$ (the value where the broken phase minimum
  appears at the critical temperature in the absence of magnetic field) for
  the same temperatures as in Fig.~\ref{fig7} and a fixed value of the
  magnetic field parametrized as $B=b\times (100\ {\mbox{GeV}})^2$, with
  $b=0.01$. The gauge parameter is $\xi=0.1$. The insert shows
  the difference ($\Delta V_{\mbox{\it eff}}$) with respect to the effective
  potential shown in Fig.~\ref{fig7}
  in the absence of magnetic field over a small region
in the $\nu$ range where the second minimum would be developing and
is now delayed by the presence of magnetic field.}
\label{fig8}
\end{figure}

\begin{figure}[t!] 
{\centering
{\epsfig{file=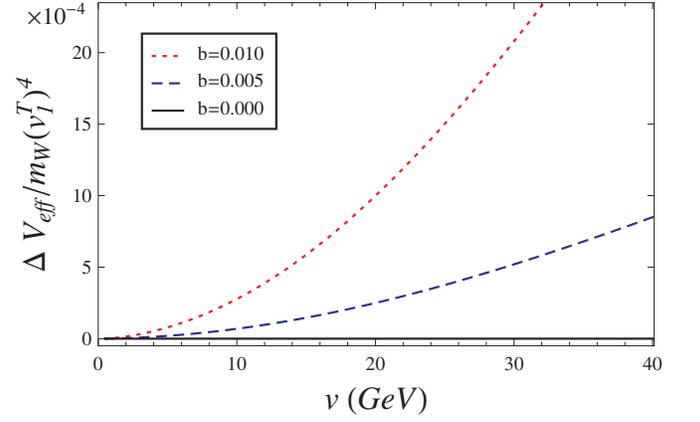, width=1\columnwidth}}
\par}
\caption{Difference between the effective potentials in the presence
and in the absence of magnetic field divided by the
fourth power of the $v$-dependent $W$ mass evaluated at $v_1^T$
(the value where the broken phase minimum appears at the critical
temperature $T=T_c^{B=0}=139.758$ GeV in the absence
  of magnetic field) for different values of the magnetic field parametrized
  as $B=b\times (100\ {\mbox{GeV}})^2$. The gauge parameter is $\xi=0.1$.
  Even though the magnetic field is weak, we can appreciate
  that the phase transition is delayed when increasing the
  values of the magnetic field.}
\label{fig9}
\end{figure}

Figure~\ref{fig8} shows the effective potential divided by
$[m_W(v_1^T)]^4=(10.5\ {\mbox{GeV}})^4$ for the same temperatures as in
Fig.~\ref{fig7} and a fixed value of the magnetic field parametrized as
$B=b\times (100\ {\mbox{GeV}})^2$, with $b=0.01$. An insert in this plot
shows the difference ($\Delta V_{\mbox{\it \textrm{eff}}}$) with respect to the effective
potential shown in Fig.~\ref{fig7} in the absence of magnetic field over a small region
in the $\nu$ range where the second minimum would be developing and
is now delayed by the presence of magnetic field. This same effect can be
observed if we keep the temperature fixed and increase the value of the
magnetic field. Given that we use a weak magnetic field, the effect is small.
In order to appreciate such effect, in Fig.~\ref{fig9} we show
the difference between the effective potentials in the presence
and in the absence of magnetic field. Starting
from zero magnetic field, for which the phase transition happens at the
critical temperature $T=139.758$ GeV, the phase transition is delayed by
increasing the values of the magnetic field, also parametrized as $B=b\times
(100\ {\mbox{GeV}})^2$, while the temperature is maintained fixed. In both
Figs.~\ref{fig8} and~\ref{fig9}, the value of the gauge parameter is $\xi=0.1$.

\begin{figure}[b!] 
{\centering
{\epsfig{file=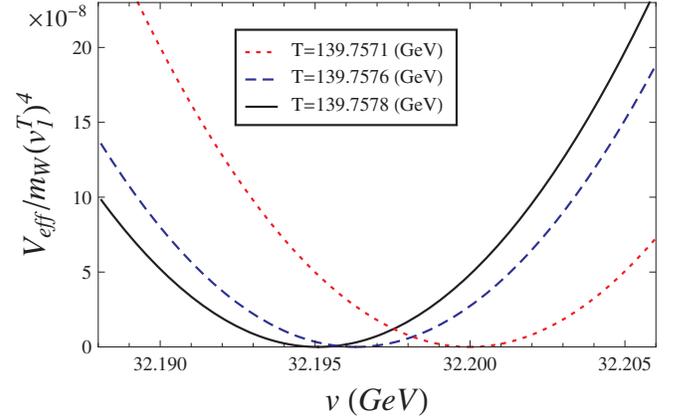, width=1\columnwidth}}
\par}
\caption{A small region around the second minima that develops in the
effective potential divided by the fourth power of the $v$-dependent
  $W$ mass evaluated at $v_1^T$ (the value where the broken phase minimum
  appears at the critical temperature in the absence
  of magnetic field) for three different values of the magnetic field
  at their corresponding critical temperatures and for $\xi=0.1$. Notice
  how for increasing
  values of the magnetic field, the phase transition starts at a lower
  temperature and the broken phase minimum is also shifted to higher
  values in such a way that the ratio $v_0/T_c$ increases at the phase
  transition.}
\label{fig10}
\end{figure}

Figure~\ref{fig10} shows a small region around the second minima that develops
in the effective potential divided by the fourth power of
the $v$-dependent $W$ mass evaluated at $v_1^T$, for three different
values of the magnetic field and computed at their corresponding critical
temperatures, keeping $\xi=0.1$. Notice that for increasing values of the
magnetic field, the phase transition starts at a lower critical temperature
$T_c$ and the broken
phase minimum $v_0$ is also shifted to higher values in such a way that
the ratio $v_0/T_c$ increases at the phase transition. This increase,
although modest, is a desired feature that can eventually help a possible
baryon asymmetry to not be washed out after the completion of the phase
transition~\cite{Kajantie1}.

\begin{figure}[t!] 
\vspace{0.5cm}
{\centering
{\epsfig{file=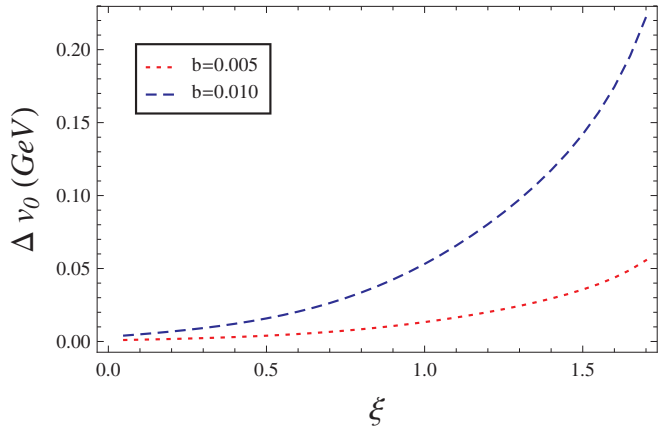, width=1\columnwidth}}
\par}
\caption{Difference of gauge parameter dependence of $v_0$ with and without
magnetic field.  The position of the broken phase
  minimum at the critical temperature, for two values of the magnetic field
  parametrized as $B=b\times (100\ {\mbox{GeV}})^2$ remains stable throughout
  a large range of $\xi$, even in the presence of magnetic field.}
\label{fig11}
\end{figure}

\begin{figure}[b!] 
\vspace{0.5cm}
{\centering
{\epsfig{file=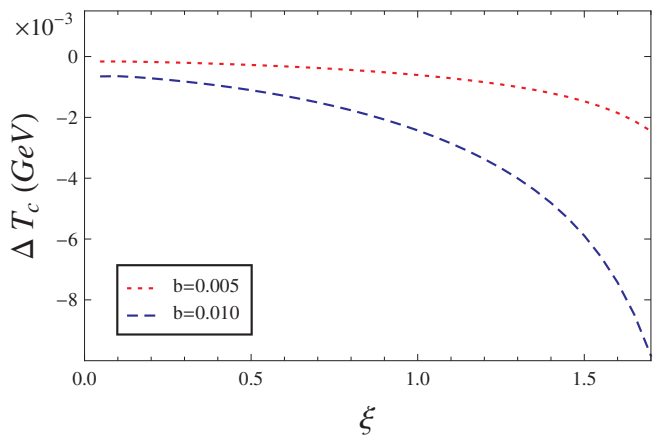, width=1\columnwidth}}
\par}
\caption{Difference of gauge parameter dependence of $T_c$ with and without
magnetic field. The position of the critical temperature for the
  development of the broken phase minimum, for two values of the magnetic field
  parametrized as $B=b\times (100\ {\mbox{GeV}})^2$ remains stable throughout
  a large range of $\xi$.}
\label{fig12}
\end{figure}

\begin{figure}[t!] 
\vspace{0.5cm}
{\centering
{\epsfig{file=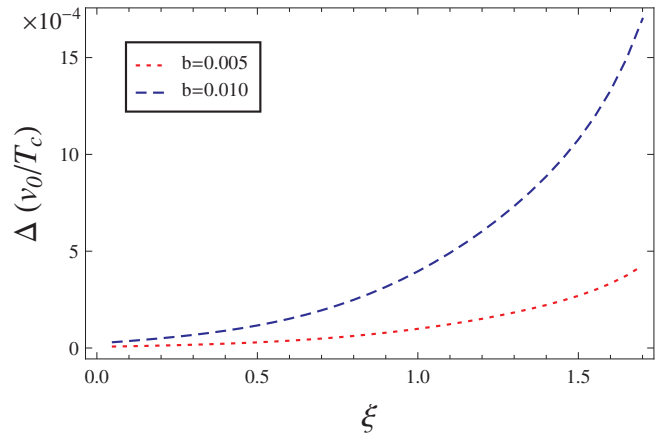, width=1\columnwidth}}
\par}
\caption{Difference of gauge parameter dependence of the ratio $v_0/T_c$ with and
without magnetic field, for two values of
  the magnetic field parametrized as $B=b\times (100\ {\mbox{GeV}})^2$. Notice
 how the value of $v_0/T_c$ is stable for a variation of $\xi$ up to 1.5. The presence
of the magnetic field does not introduce a strong gauge dependence. }
\label{fig13}
\end{figure}

To explore the gauge parameter dependence of the effective potential's
relevant parameters, Figs.~\ref{fig11} --~\ref{fig13} show the difference in
the behavior with and without magnetic field of
$v_0$, $T_c$ and the ratio $v_0/T_c$ for values of $\xi$ around 1. We use
a fixed value of the magnetic field, parametrized as
$B=b\times (100\ {\mbox{GeV}})^2$, with $b=0,\ 0.005,\ 0.01$.
We can see that $v_0$, $T_c$ and $v_0/T_c$ remain stable for a variation of $\xi$ up to 1.5. For these small values of the magnetic field strenght, consistent with the assumed hierarchy of energy scales,
the magnetic field does not introduce a strong gauge dependence.

\section{Discussion and Conclusions}\label{VII}

In conclusion, we have studied the symmetry restoration problem in the MSM at
finite temperature in the presence of an external magnetic field. The study
has been carried out by analyzing the finite temperature effective
potential. We have shown that
in  the presence of the magnetic field, the EWPT has become a stronger
first order phase transition. Our treatment has been implemented for the case
of weak magnetic fields
for the hierarchy of scales $ e B \ll m^2 \ll T^2$ where $m$ is taken as a
generic mass involved in the calculation.

We have explicitly worked with the degrees of freedom in the
broken symmetry phase where the external magnetic field belongs to the
$U(1)_{em}$ group and in a covariant gauge, with an arbitrary value of a
single gauge fixing parameter. The calculation is carried out up to the
contribution of ring diagrams. To include the effects of the magnetic field,
we have made use of the Schwinger proper-time method to describe the
particle's propagators. In this way, the contribution from all
Landau levels has been accounted for. We have carried out a systematic
expansion up to order $(e B)^2$ and have computed, as an intermediate step,
the magnetic field and gauge fixing parameter dependent self-energies in the
MSM~\cite{tesisjorge}.

The presence of the external magnetic field gives rise to terms in the
effective potential proportional to $1/\tilde{m}_i^3$, where
$\tilde{m}_i^2=m_i^2+\Pi_1$, coming from tadpole diagrams in the boson
self-energies where the loop particle is a charged scalar and $m_i$ are their
masses. These terms are the relevant ones for the strengthening of the order
of the phase transition. The results are in qualitative and quantitative
agreement with the ones previously found~\cite{sap} using symmetry restored
degrees of freedom. The ratio of the broken phase vacuum expectation value to
the critical temperature at the phase transition increases with increasing
values of the magnetic field. This feature works in favor of the suppression
of the sphaleron induced transitions in the broken phase after the EWPT is
completed and thus against a washing out of a possibly produced baryon
number. The increase in this ratio is modest, though it is important to notice
that we have used very restrictive values of the field strength, consistent
with the working hierarchy of energy scales used in this work. The leading
temperature terms of the effective potential are gauge parameter independent,
although the subleading ones, i.e., the ones coming with the magnetic field
turn out to be gauge parameter dependent. The dependence on this parameter of
phase transition observables such as the position of the minimum in the
symmetry broken phase and the critical temperature for the transition is not
negligible for values of $\xi$ near $\xi\sim 1$. This dependence signals
that the truncation of the magnetic field power series might not be a gauge
invariant procedure. This point deserves a closer look and is being
investigated~\cite{jorge}.

\begin{widetext}

\section*{Acknowledgments}

Support has been received in part by DGAPA-UNAM under PAPIIT Grants
Nos. IN116008 and IN112308 and by CONACyT-M\'exico under Grant
52547-F. J. Navarro acknowledges support received by  Universidad del
Atl\'antico. M.E.T.-Y. thanks the \textit{Programa de Intercambio
UNAM-UNISON} for support. A.A. and J.N. thank the support provided by
CONACyT-M\'exico under Grant 40025-f.

\section*{Appendix A: Scalar self-energies in arbitrary gauge}

In this appendix, we list the results for the
self-energy diagrams depicted in Fig.~\ref{fig1} (for the neutral Goldstone
boson there is no contribution of the type shown in diagram
$(c)$ of Fig.~\ref{fig1}).

\subsection{Higgs boson ($H$) self-energy in arbitrary gauge}

\begin{eqnarray}
\label{piHa}
 \Pi _a^H &=&
\lambda \frac{T^2  }{2}
\left[ 1 - \frac{1}{2\pi T} \left(2 m_1 + m_3 + 3  m_4
  \right)\right]
-\lambda \frac{(eB)^2 }{48 \pi m_1^2} \left[ \frac{T}{m_1}
+\frac{\zeta(3) m_1^2}{4 \pi^3 T^2} \right]  \\ \nonumber \\
\label{piHb}
\Pi _b^H  &=& \frac{T^2}{16} \left[g^2(3+\xi)+g'^2\left(1+\frac{\xi }{3}
  \right) - \frac{3 + \xi ^{3/2}}{\pi T} \left(2 g^2 m_W
+ (g^2 + g'^2) m_Z \right)\right] \nonumber \\
&&  + \frac{(eB)^2 g^2}{64 \pi  m_W^2} \left[
\left(3+\frac{35}{3 \xi^{1/2}}\right) \frac{T}{m_W}+\frac{11 \zeta(3)
  m_W^2}{3 \pi ^3 T^2} \right]
\\ \nonumber \\
\label{piHc}
\Pi _c^H  &=&
-\frac{\left(3 g^4+2 g^2 g'^2+g'^4\right)  m_W^2}{4 g^2 \pi ^2}
\ln\left(\frac{m_W}{T}\right) \\ \nonumber \\
&& -\frac{T}{2\pi}
\left[g^2 m_W  + \frac{g^2 m_W^2}{2 m_Z}
+ \frac{g'^2 m_W^2}{m_Z}
+\frac{g'^4 m_W^2}{2 g^2 m_Z} \right]
\left(\frac{3 }{4
 }+\frac{5  \xi ^{3/2}}{4  }-\frac{ \xi ^{3/2}}{
 \left(1+\xi^{1/2}\right)}-\frac{ \xi ^2 }{
  \left(1+\xi^{1/2}\right)}\right)
 \nonumber \\
&& +(eB)^2 \frac{g^2 T}{\pi m_W^3} \left(-\frac{83 }{192
}+\frac{13 }{12
  \left(1+\xi^{1/2}\right)^4  }+\frac{13 }{48
  \left(1+\xi^{1/2}\right)^3  } \right. \nonumber \\
&& \left. -\frac{5}{4  \left(1+\xi^{1/2}\right)^2  }-\frac{1}{192
  \xi^{1/2}  }+\frac{5}{24  \left(1+\xi^{1/2}\right)^4 \xi^{1/2}
} -\frac{7 }{16  \left(1+\xi^{1/2}\right)^3 \xi^{1/2}  }\right. \nonumber \\
&& \left.-\frac{1}{2
   \left(1+\xi^{1/2}\right)^2 \xi^{1/2}  }+\frac{9
 \xi^{1/2} }{4  \left(1+\xi^{1/2}\right)^4  }
 +\frac{3 \xi^{1/2} }{16
  \left(1+\xi^{1/2}\right)^3  }-\frac{3 \xi^{1/2} }{4
  \left(1+\xi^{1/2}\right)^2  }
 \right. \nonumber \\
&& \left.+\frac{9
\xi }{4   \left(1+\xi^{1/2}\right)^4  }
-\frac{17 \xi }{48   \left(1+\xi^{1/2}\right)^3  }+\frac{7 \xi
  ^{3/2} }{8  \left(1+\xi^{1/2}\right)^4  }+\frac{1}{4 \left( 1 +
  \xi^{1/2} \right)}\right)
-\frac{9 (eB)^2 g^2 m_W^2 \zeta(5)}{256 \pi ^6 T^4}
\end{eqnarray}

\begin{eqnarray}
\label{piHd}
\Pi _d^H &=& -T^2 \frac{(3g^2+g'^2)\xi}{48}
\nonumber \\
&& +T \left( \frac{g^2 m_1^3}{8 \pi m_W(  m_1+
  m_W)} \left(\frac{m_1^2(  m_1+  m_W)}{
  m_W \left(m_1^2-\xi m_W^2\right)} -\frac{ \xi ^{5/2}
  m_W^4(  m_1+  m_W)}{
  m_1^3\left(m_1^2-\xi
  m_W^2\right)}-\frac{m_1}{m_W}
-1\right)\right.\nonumber \\
&&\left.+(g^2 + g'^2)\frac{m_3^3}{16 \pi  m_Z(  m_3+
  m_Z)} \left(
\frac{ m_3^2(  m_3+  m_Z)}{m_Z
  \left(m_3^2-\xi  m_Z^2\right)}
-\frac{ \xi ^{5/2} m_Z^4(  m_3+  m_Z)}{ m_3^3
  \left(m_3^2-\xi  m_Z^2\right)}
-\frac{m_3}{m_Z }
-1 \right) \right. \nonumber \\
&& +(eB)^2 \frac{g^2 T}{\pi}  \left(  \frac{1}{ (m_1+m_W)^4} \left(
-\frac{9  m_1}{16
 } -\frac{7  m_1^3}{32   m_W^2 }-\frac{9
 m_1^2}{16  m_W}-\frac{13  m_W}{48 }-\frac{5
 m_W^2}{96   m_1} \right) \right.\nonumber \\
&&\left. +\frac{1}{ (m_1+m_W)^3} \left( \frac{31
}{192}+\frac{11
 m_1^2}{64   m_W^2 } +\frac{13  m_1}{64
  m_W}+\frac{3  m_W}{64  m_1 }
 \right)
+ \frac{1}{ m_W (m_1+m_W)^2} \left( \frac{
m_1}{24  m_W }+\frac{1}{48 } \right) \right. \nonumber \\
&&\left. + \frac{m_1}{ \left(m_1+\xi^{1/2}
   m_W\right)^4} \left(\frac{49
 \xi  }{96 }+\frac{7  m_1^2}{32 m_W^2 }
+\frac{49  \xi^{1/2} m_1}{96  m_W
}+\frac{7  \xi ^{3/2} m_W}{32 m_1} \right)\right.\nonumber \\
&&\left.  - \frac{1}{ \left(m_1+\xi^{1/2} m_W\right)^3}
\left(\frac{11  \xi }{64}+\frac{11  m_1^2}{64  m_W^2 }
+\frac{9  \xi^{1/2} m_1}{32   m_W}\right) \right.\nonumber \\
&&\left.- \frac{1}{ m_W \left(m_1+\xi^{1/2}
  m_W\right)^2} \left(\frac{ m_1}{24   m_W}
+\frac{ \xi^{1/2}}{48   }\right) \right)
-\frac{7 (eB)^2 g^2 \zeta(3)}{384 \pi ^4 T^2} \\ \nonumber \\
\label{piHe}
\Pi _e^H &=& -\frac{\lambda ^2 m_W^2}{g^2 \pi} \left[ \frac{12}{ \pi}  \ln
  \left(\frac{m_W}{T}\right) +T \left(\frac{2 }{
    m_1}+\frac{1}{ m_3}+\frac{9 }{  m_4}\right)
  \right] +\frac{(eB)^2\lambda ^2 m_W^2}{4 g^2 \pi m_1^4}\left[\frac{T
  }{m_1}+\frac{\zeta(5)m_1^4}{16 \pi ^5 T^4}\right] \\ \nonumber \\
\label{piHf}
 \Pi _f^H  &=& \frac{m_t^2 T^2}{2 v_0^2} \left[1 +  \frac{3 m_t^2 }{ \pi ^2
     T^2} \ln\left(\frac{4 m_W}{T}\right) \right] -\frac{(eB)^2
   m_t^2}{ 2 \pi ^4 T^2 v_0^2} \left[\frac{7 \zeta(3)}{9}  - \frac{31 m_t^2
     \zeta(5)}{16 \pi ^2 T^2}\right] \\ \nonumber \\
\label{piHg}
\Pi _g^H  &=& \frac{g^2 \xi ^2 m_W^2}{16 \pi} \left[
\frac{1}{2 \pi} \left(3+2 \frac{g'^2}{g^2}+\frac{g'^4}{g^4}\right)
\ln\left(\frac{m_W}{T}\right) +
\frac{T}{m_W}
+ \left(1+2 \frac{g'^2}{g^2}+\frac{g'^4}{ g^4}\right)\frac{T}{2
  m_W}
\right] \nonumber \\
&& -\frac{(eB)^2 g^2 \xi ^2 m_W^2}{128 \pi  m_W^4} \left[
  \frac{T}{m_W} + \frac{\zeta(5)m_W^4}{16 \pi
^5 T^4}\right]
\end{eqnarray}

\subsection*{Neutral Goldstone boson ($G^0$) self-energy in arbitrary gauge}

\begin{eqnarray}
\label{piG0a}
 \Pi _a^{G^0}&=&
\frac{\lambda T^2}{2} \left[ 1 - \frac{1}{2 \pi T} \left(2 m_1 + 3
  m_3 + m_4 \right)-\frac{(eB)^2  }{24 \pi T
    m_1^3}-\frac{(eB)^2   \zeta(3)}{96 \pi ^4 T^4} \right] \\ \nonumber \\
\label{piG0b}
\Pi _b^{G^0} &=&
\frac{T^2}{16} \left[ (3+\xi )g^2 + \left(1+ \frac{\xi }{3}\right)g'^2
  \right]
-\frac{T}{16\pi} \left[ (6+2\xi ^{3/2})g^2 m_W + (g^2+g'^2)(3+\xi
  ^{3/2}) m_Z\right]\nonumber \\
&& +(eB)^2 T \frac{g^2}{\pi m_W^3}
\left(\frac{3}{64 }+\frac{35}{192 \xi^{1/2}} +\frac{11
\zeta(3) m_W^3}{192 \pi ^3 T^3} \right) \\ \nonumber \\
\label{piG0d}
\Pi _d^{G^0} &=&
-\left(3 g^2+g'^2\right)\frac{\xi T^2}{48}
-\frac{g^2 T}{8\pi} \left(
\left(1+\frac{m_1}{m_W}\right)\frac{ m_1^3}{
  m_W ( m_1+ m_W)}
 \right. \nonumber \\
&&\left. +\frac{1}{ \left(m_1^2-\xi  m_W^2\right)} \left(\xi
^{5/2} m_W^3
-\frac{ m_1^5}{m_W^2} \right)
+\left(1+\frac{g'^2}{g^2}\right)\frac{m_4^4}{2
  m_Z^2 (m_4+ m_Z)} \right) \nonumber \\
&& +(eB)^2 \frac{g^2 T}{\pi} \left(
\frac{1}{ (m_1+m_W)^4}\left(
-\frac{9   m_1}{16}-\frac{7   m_1^3}{32  m_W^2}
-\frac{9  m_1^2}{16 m_W}-\frac{13   m_W}{48 }-\frac{5
  m_W^2}{96 } \right) \right. \nonumber
\end{eqnarray}

\begin{eqnarray}
&&\left.
+\frac{1 }{  (m_1+m_W)^3} \left(
\frac{31  }{192 }+\frac{11 m_1^2}{64  m_W^2 }+\frac{13
  m_1}{64  m_W}
+\frac{3   m_W}{64 m_1 } \right)
 + \frac{1}{ m_W (m_1+m_W)^2} \left(\frac{
m_1}{24 m_W}+\frac{ 1}{48  } \right)\right. \nonumber \\
&&\left.+
\frac{m_1}{  \left(m_1+\xi^{1/2} m_W\right)^4}\left(
\frac{49\xi }{96 }+\frac{7   m_1^2}{32  m_W^2 }+\frac{49
  \xi^{1/2} m_1}{96  m_W}+\frac{7   \xi ^{3/2} m_W}{32
  m_1}\right)\right. \nonumber \\
&&\left.+\frac{1}{  \left(m_1+\xi^{1/2}
  m_W\right)^3}\left(-\frac{11   \xi }{64}
-\frac{11   m_1^2}{64  m_W^2}-\frac{9   \xi^{1/2}
  m_1}{32  m_W} \right) \right. \nonumber \\
&&\left. +\frac{1}{  m_W \left(m_1+\xi^{1/2}
  m_W\right)^2}\left(-\frac{  m_1}{24  m_W}-\frac{
  \xi^{1/2}}{48 }\right) \right)  \\ \nonumber \\
\label{piG0e}
 \Pi _e^{G^0} &=&
- \frac{\lambda ^2  m_W^2}{g^2} \left[ \frac{2 }{\pi
    ^2}\ln\left(\frac{m_W}{T}\right)+\frac{16 T }{4\pi (  m_3+
    m_4)} \right] \\ \nonumber \\
\label{piG0f}
 \Pi _f^{G^0} &=&
\frac{m_t^2}{2 v_0^2} \left[T^2-\frac{3 m_t^2 }{ \pi ^2}\ln\left(\frac{4
    m_W}{T}\right)-\frac{7
(eB)^2  \zeta(3)}{9 \pi ^4 T^2}+\frac{31 (eB)^2 m_t^2 \zeta(5)}{16 \pi ^6 T^4}
  \right] \\ \nonumber \\
\label{piG0g}
 \Pi _g^{G^0} &=& -\frac{g^2 \xi ^2  m_W^2}{16\pi}\left[
\frac{1}{\pi}\ln\left(\frac{m_W}{T}\right)+\frac{ T }{  m_W}-\frac{(eB)^2 T
}{8  m_W^5}-\frac{(eB)^2
 \zeta(5)}{128 \pi ^5 T^4} \right]
\end{eqnarray}

\subsection*{Charged Goldstone boson ($G^c$) self-energy in arbitrary gauge}
\begin{eqnarray}
\label{piGCa}
\Pi _a^{G^c}&=&
\frac{T^2 \lambda }{2}\left[ 1 - \frac{1}{2\pi T} \left(2m_1
  +m_3+m_4\right)-\frac{(eB)^2 }{12 \pi T
    m_1^3}-\frac{(eB)^2   \zeta(3)}{48 \pi ^4 T^4} \right] \\ \nonumber \\
\label{piGCb}
 \Pi _b^{G^c} &=&
T^2 \frac{(\xi + 3)( 3g^4 + g'^4 +4 g^2 g'^2 )}{48\left(g^2+g'^2\right)}
\nonumber \\
&& -T \frac{(\xi ^{3/2}+3)}{4 \pi \left(g^2+g'^2\right)}\left(
\frac{g^2 \left(g^2+g'^2\right) m_W}{2  }
+ \frac{\left(g^2 - g'^2\right)^2 m_Z}{4}
+ g^2 g'^2 m_\gamma \right) \nonumber \\
&& +(eB)^2 T \frac{g^2}{\pi m_W^3}
\left(\frac{3}{64 }+\frac{35}{192  \xi^{1/2}}\right)+\frac{11 (eB)^2 g^2
  \zeta(3)}{192 \pi ^4 T^2}  \\ \nonumber \\
\label{piGCc}
\Pi _c^{G^c} &=&
-\frac{g'^2  m_W^2}{2 \pi ^2} \ln\left(\frac{m_W}{T}\right)\nonumber \\
&& +\frac{T m_W^2}{4\pi \left(g^2+g'^2\right)} \left(-\frac{3 g'^4 }{ (
  m_W+ m_Z)}
- \frac{g'^4 }{ \left(\xi^{1/2} m_W+ m_Z\right)} \left(\xi
+\frac{\xi^{1/2}   m_Z}{   m_W} +\frac{ m_Z^2}{
  m_W^2 } \right) \right. \nonumber \\
&&\left. -\frac{g'^4  }{  \left(  m_W+ \xi^{1/2}
  m_Z\right)}\left(\xi
+\frac{ m_W^2}{  m_Z^2 }
+\frac{\xi^{1/2}   m_W}{ m_Z } \right)
+\frac{g'^4 }{ \left(\xi  m_W^2-m_Z^2\right)}\left(\frac{\xi
  ^{5/2}   m_W^3}{  m_Z^2}
-\frac{m_Z^3}{   m_W^2 } \right) \right. \nonumber \\
&& \left.+ \frac{g'^4}{\left(m_W^2-\xi  m_Z^2\right)}\left(
\frac{m_W^3}{   m_Z^2 }
-\frac{\xi ^{5/2}   m_Z^3}{   m_W^2 }\right)
-\frac{g'^4\xi ^{3/2} }{ \left( m_W^2-  m_Z^2\right)}\left(
\frac{  m_W^3}{  m_Z^2 }
+\frac{  m_Z^3}{ m_W^2 } \right) \right. \nonumber \\
&& \left.-\frac{3 g^2 g'^2  }{   (  m_W+  m_\gamma)}
-\frac{g^2 g'^2  }{   \left(  \xi^{1/2} m_W+
  m_\gamma\right)}\left(
\xi
+\frac{\xi^{1/2}   m_\gamma}{   m_W }
+\frac{  m_\gamma^2}{  m_W^2 } \right) \right. \nonumber \\
&& \left. -\frac{g^2 g'^2   }{ \left(  m_W+  \xi^{1/2}
  m_\gamma\right)} \left(
\xi
+\frac{m_W^2}{ m_\gamma^2}
+\frac{\xi^{1/2} m_W}{   m_\gamma } \right)
+ \frac{g^2 g'^2 }{ \left(\xi  m_W^2-m_\gamma^2\right)} \left(
\frac{\xi ^{5/2}   m_W^3}{   m_{\gamma}^2 }
-\frac{ m_\gamma^3}{m_W^2 } \right) \right. \nonumber \\
&& \left. +\frac{g^2 g'^2 m_W^3}{   m_\gamma^2
  \left(m_W^2-\xi  m_{\gamma}^2\right)} -\frac{g^2 g'^2 \xi ^{5/2}
  m_{\gamma}^3}{   m_W^2 \left(m_W^2-\xi
  m_\gamma^2\right)}
-\frac{g^2 g'^2 \xi ^{3/2} }{   \left(m_W^2-
  m_\gamma^2\right)} \left(
\frac{m_W^3}{m_{\gamma}^2}
-\frac{m_\gamma^3}{m_W^2}\right)\right) \nonumber
\end{eqnarray}
\normalsize

\begin{eqnarray}
&&+(eB)^2 T \frac{m_W^2 }{\pi \left(g^2+g'^2\right)} \left(
-\frac{g'^4   }{ 16(m_W+m_Z)^4}\left(
\frac{13 }{m_W }
+\frac{2 m_W}{  m_Z^2 }
+\frac{8  }{  m_Z }
+\frac{12  m_Z}{ m_W^2 }
+\frac{3  m_Z^2}{ m_W^3 } \right) \right. \nonumber \\
&& \left.
-\frac{g'^4}{(m_W+m_Z)^3}\left(
\frac{77  }{96  m_W^2 }
-\frac{1  }{24  m_Z^2 }
-\frac{1  }{8   m_W m_Z}
+\frac{5 m_Z}{32  m_W^3 } \right) \right. \nonumber \\
&& \left.
+\frac{g'^4  }{ (m_W+m_Z)^2}\left(
\frac{1 }{4 m_W^3 }
+\frac{1}{24  m_W m_Z^2 }
+\frac{1}{12  m_W^2 m_Z } \right) \right. \nonumber \\
&& \left.
+\frac{5 g'^4}{ \left(\xi^{1/2} m_W+m_Z\right)^4}\left(
\frac{ \xi^{1/2}  }{8  m_W }
+\frac{ \xi ^{3/2}   m_W}{48  m_Z^2 }
+\frac{ \xi   }{12   m_Z }
+\frac{ m_Z}{12 m_W^2 }
+\frac{ m_Z^2}{48 \xi^{1/2} m_W^3 } \right) \right. \nonumber \\
&& \left.
-\frac{g'^4  }{\left(\xi^{1/2} m_W+m_Z\right)^3}\left(
\frac{31  }{96   m_W^2 }
+\frac{3 \xi   }{32  m_Z^2 }
+\frac{9 \xi^{1/2}  }{32 m_W m_Z }
+\frac{7  m_Z}{32  \xi^{1/2} m_W^3 }
\right) \right. \nonumber \\
&& \left.
-\frac{g'^4 }{\left(\xi^{1/2} m_W+m_Z\right)^2}\left(
\frac{1}{4  \xi^{1/2} m_W^3 }
+\frac{\xi^{1/2}  }{24 m_W m_Z^2 }
+\frac{1}{12 m_W^2 m_Z } \right) \right. \nonumber \\
&& \left.
+\frac{g'^4 }{ \left(m_W+\xi^{1/2} m_Z\right)^4}\left(
\frac{17\xi   }{24  m_W }
+\frac{ m_W}{8  m_Z^2 }
+\frac{\xi^{1/2}  }{2  m_Z }
+\frac{ \xi ^{3/2}   m_Z}{3  m_W^2 } \right) \right. \nonumber
\\
&&\left.
-\frac{g'^4}{\left(m_W+\xi^{1/2} m_Z\right)^3}\left(
\frac{\xi   }{4 m_W^2 }
+\frac{1}{24 m_Z^2 }
+\frac{\xi^{1/2}  }{8 m_W m_Z } \right)
-\frac{g'^4 }{ \left(m_W+\xi^{1/2} m_Z\right)^2}\left(
\frac{1}{24 m_W m_Z^2 }
+\frac{\xi^{1/2}  }{12 m_W^2 m_Z } \right)\right. \nonumber \\
&& \left.
-\frac{g'^4 }{\xi ^{1/2}  \left( m_W+ m_Z\right)^4}\left(
\frac{29   }{48     m_W }
+\frac{5   m_W}{48    m_Z^2}
+\frac{5   }{12  m_Z }
+\frac{  m_Z}{3  m_W^2 }
\right) \right. \nonumber \\
&& \left.
+\frac{g'^4   }{\xi^{1/2}   \left( m_W+m_Z\right)^3} \left(
\frac{1   }{4      m_W^2}
+\frac{3   }{32    m_Z^2 }
+\frac{9   }{32     m_W m_Z } \right)
+\frac{g'^4   }{12 \xi^{1/2} \left(m_W+ m_Z\right)^2}\left(
\frac{1  }{2  m_W m_Z^2 }
+\frac{1  }{  m_W^2 m_Z } \right)
\right. \nonumber \\
&&\left.
-\frac{g^2 g'^2  }{  (m_W+m_\gamma)^4} \left(
\frac{13  }{16  m_W }
+\frac{  m_W}{8  m_\gamma^2 }
+\frac{ 1}{2  m_\gamma }
+\frac{3   m_\gamma}{4    m_W^2 }
+\frac{3   m_\gamma^2}{16 m_W^3} \right)
 \right. \nonumber \\
&& \left.
-\frac{g^2 g'2}{ m_W^2(m_W+m_\gamma)^3} \left(
\frac{77}{96 m_W^2}
-\frac{1}{24 m_{\gamma}^2 }
-\frac{1}{8 m_W m_\gamma }
+\frac{5 m_{\gamma}}{32  m_W^3 } \right)
\right. \nonumber \\
&& \left.
+\frac{g^2 g'^2  }{ m_W (m_W+m_\gamma)^2}\left(
\frac{1 }{4  m_W^2}
+\frac{1}{24  m_\gamma^2 }
+\frac{1}{12    m_W m_\gamma } \right)
\right. \nonumber \\
&& \left.
+\frac{5 g^2 g'^2}{ \left(\xi^{1/2} m_W+m_\gamma\right)^4}
\left(
\frac{\xi ^{3/2}   m_W}{ m_{\gamma}^2 }
+\frac{\xi   }{12  m_\gamma }
+\frac{m_{\gamma}}{12      m_W^2 }
+\frac{m_\gamma^2}{48      \xi^{1/2} m_W^3 }
+\frac{ \xi^{1/2}  }{8   m_W }
 \right)
\right. \nonumber \\
&& \left.
-\frac{g^2 g'^2  }{ \left(\xi^{1/2}
  m_W+m_\gamma\right)^3}\left(
\frac{31  }{96 m_W^2 }
+\frac{3  \xi }{32  m_\gamma^2 }
+\frac{9  \xi^{1/2}  }{32   m_W m_\gamma }
+\frac{7    m_\gamma}{32  \xi^{1/2} m_W^3 } \right)
\right. \nonumber \\
&& \left.
-\frac{g^2 g'^2  }{\left(\xi^{1/2}
  m_W+m_\gamma\right)^2}\left(
\frac{1  }{4   \xi^{1/2} m_W^3 }
+\frac{\xi^{1/2}  }{24  m_W m_\gamma^2 }
+\frac{1  }{12  m_W^2 m_\gamma } \right)
\right. \nonumber \\
&& \left.
+\frac{g^2 g'^2 }{ \left(m_W+\xi^{1/2} m_\gamma\right)^4}
\left(
\frac{17  \xi   }{24  m_W }
+\frac{  m_W}{8  m_{\gamma}^2 }
+\frac{\xi^{1/2}  }{2  m_\gamma }
+\frac{\xi ^{3/2}  m_\gamma}{3  m_W^2 }
\right)
\right. \nonumber \\
&& \left.
-\frac{g^2g'^2  }{ \left(m_W+\xi^{1/2}
  m_\gamma\right)^3}\left(
\frac{\xi   }{4  m_W^2 }
+\frac{\xi^{1/2}  }{8  m_W m_{\gamma} }
+\frac{1  }{24      m_\gamma^2 }
\right)
-\frac{g^2 g'^2  }{24  m_W m_\gamma
  \left(m_W+\xi^{1/2} m_\gamma\right)^2}\left(
\frac{1  }{    m_\gamma }
+\frac{ 2 \xi^{1/2}}{   m_W }
\right)
\right. \nonumber \\
&& \left.
-\frac{g^2 g'^2}{\xi ^{1/2} \left(m_W+m_\gamma\right)^4}
\left(
\frac{29   }{48 m_W }
+\frac{5    m_W}{48 m_\gamma^2 }
+\frac{   m_\gamma}{3 m_W^2}
+ \frac{5}{ 12  m_\gamma}\right)
\right. \nonumber \\
&& \left.
+\frac{g^2 g'^2  }{\xi^{1/2}\left(
  m_W+m_\gamma\right)^3}\left(
\frac{1 }{4  m_W^2 }
+\frac{3 }{32 m_\gamma^2 }
+\frac{9 }{32  m_W m_\gamma } \right)
+\frac{g^2 g'^2 }{  \xi^{1/2} \left( m_W+
  m_\gamma\right)^2}\left(
\frac{1  }{24  m_W m_\gamma^2 }
+\frac{1 }{12   m_W^2 m_\gamma }\right) \right) \nonumber \\
&&
-\frac{(eB)^2}{T^4} \frac{7 g'^2 m_W^2 \zeta(5)}{512 \pi ^6} \nonumber \\
\end{eqnarray}

\begin{eqnarray}
\label{piGCd}
 \Pi _d^{G^c} &=&
-\frac{\xi \left(3g^4+4g^2 g'^2+g'^4\right) T^2}{48\left(g^2+g'^2\right)}
\nonumber \\
&& +T \left(
-\frac{g^2 m_3^3}{16 \pi m_W^2}
-\frac{g^2 m_4^3}{16 \pi m_W^2 }
-\frac{(g^2-g'^2)^2 m_1^3}{16\pi m_Z^2}
-\frac{g^2 g'^2 m_1^3}{4 \pi m_\gamma^2}
\right. \nonumber \\
&& \left.
+\frac{g^2}{16 \pi \left(m_3^2-\xi m_W^2\right)}\left(
\frac{m_3^5}{ m_W^2 } -\xi ^{5/2} m_W^3 \right)
+\frac{g^2}{16 \pi \left(m_4^2-\xi  m_W^2\right)} \left(
\frac{m_4^5}{m_W^2} - \xi ^{5/2} m_W^3\right)
\right. \nonumber \\
&& \left.
+\frac{(g^2-g'^2)^2}{16 \pi \left(m_1^2-\xi  m_Z^2\right)}\left(
\frac{m_1^5}{    m_Z^2 } -\xi ^{5/2} m_Z^3 \right)
-\frac{g^2 g'^2}{4 \pi \left(m_1^2-\xi  m_\gamma^2\right)}
\left(
\frac{m_1^5}{ m_\gamma^2 }
-\xi ^{5/2} m_\gamma^3 \right)
\right) \nonumber \\
&& + \frac{(eB)^2 T}{\pi \left(g^2+g'^2\right)} \left(
-\frac{g^2 }{  (m_3+m_W)^4}\left(
\frac{m_3}{8 }
+\frac{m_3^3}{12  m_W^2 }
+\frac{17 m_3^2}{96   m_W }
+\frac{ m_W}{32  } \right)
\right. \nonumber \\
&& \left.
+\frac{g^2}{  (m_3+m_W)^3}\left(
\frac{1}{96}
+\frac{m_3^2}{16 m_W^2 }
+\frac{m_3}{32 m_W} \right)
+\frac{g^2}{\pi  (m_4+m_W)^3}\left(
\frac{1}{96 }
+\frac{m_4^2}{16 m_W^2 }
+\frac{m_4}{32  m_W } \right)
\right. \nonumber \\
&& \left.
+\frac{g^2}{ (m_3+m_W)^2}\left(
\frac{1}{96 }
+\frac{m_3}{48 m_W} \right)
+\frac{g^2}{  m_W (m_4+m_W)^2}\left(
\frac{1}{96 }
+\frac{m_4}{48  m_W } \right)
\right. \nonumber \\
&& \left.
-\frac{g^2}{  (m_4+m_W)^4} \left(
\frac{m_4}{8 }
+\frac{m_4^3}{12 m_W^2}
+\frac{17 m_4^2}{96 m_W }
+\frac{m_W}{32 } \right)
\right. \nonumber \\
&& \left.
+\frac{g^2}{  \left(m_3+\xi^{1/2} m_W\right)^4}\left(
\frac{5 \xi  m_3}{48   }
+\frac{ m_3^3}{12  m_W^2 }
+\frac{29 \xi^{1/2} m_3^2}{192  m_W}
+\frac{5 \xi ^{3/2} m_W}{192   } \right)
\right. \nonumber \\
&& \left.
-\frac{g^2}{ \left(m_3+\xi^{1/2} m_W\right)^3}\left(
\frac{3 \xi }{128 }
+\frac{m_3^2}{16 m_W^2 }
+\frac{9 \xi^{1/2} m_3}{128   m_W} \right)
\right. \nonumber \\
&& \left.
-\frac{g^2}{ m_W \left(m_3+\xi^{1/2}
  m_W\right)^2}\left(
\frac{m_3}{48 m_W }
+\frac{\xi^{1/2}}{96   } \right)
\right. \nonumber \\
&& \left.
+\frac{g^2}{  \left(m_4+\xi^{1/2} m_W\right)^4}\left(
\frac{5 \xi  m_4}{48  }
+\frac{m_4^3}{12  m_W^2}
+\frac{29 \xi^{1/2} m_4^2}{192  m_W }
+\frac{5 \xi ^{3/2} m_W}{192 } \right)
\right. \nonumber \\
&& \left.
-\frac{g^2 }{  \left(m_4+\xi^{1/2} m_W\right)^3}\left(
\frac{3\xi }{128 }
+\frac{ m_4^2}{16  m_W^2}
+\frac{9 \xi^{1/2} m_4}{128 m_W }
\right)
\right. \nonumber \\
&& \left.
-\frac{g^2 }{  m_W \left(m_4+\xi^{1/2}
  m_W\right)^2}\left(
\frac{m_4}{48   m_W }
+\frac{\xi^{1/2}}{96  }
\right)
\right. \nonumber \\
&& \left.
-\frac{5 (g^2-g'^2)^2 m_1}{192  (m_1+m_Z)^4}\left(
6
+\frac{m_1^2}{ m_Z^2 }
+4 \frac{m_1}{  m_Z}\right)
+\frac{5 (g^2-g'^2)^2 m_Z}{192  (m_1+m_Z)^4}\left(
4
+\frac{ m_Z}{    m_1 }
 \right)
\right. \nonumber \\
&& \left.
+\frac{9(g^2-g'^2)^2}{128  (m_1+m_Z)^3}\left(
 1
+\frac{  m_1^2}{3 m_Z^2}
+\frac{  m_1}{ m_Z }
+\frac{  m_Z}{ 3m_1 }
\right)
\right. \nonumber \\
&& \left.
+\frac{(g^2-g'^2)^2 }{  \left(m_1+\xi^{1/2} m_Z\right)^4}
\left(
\frac{29 \xi  m_1}{192}
+\frac{5 m_1^3}{192  m_Z^2 }
+\frac{5 \xi^{1/2}m_1^2}{48 m_Z }
+\frac{\xi ^{3/2} m_Z}{12  }
\right)
\right. \nonumber \\
&& \left. +\frac{(g^2-g'^2)^2 }{  \left(m_1+\xi^{1/2}
    m_Z\right)^3}
\left(
-\frac{\xi}{16} - \frac{3 m_1^2}{128 m_Z^2}
-\frac{9 \xi^{1/2} m_1}{128 m_Z}
\right)
\right. \nonumber \\
&& \left.
-\frac{5 g^2 g'^2}{ (m_1+m_\gamma)^4}\left(
\frac{m_1}{8 }
+\frac{m_1^3}{48    m_\gamma^2 }
+\frac{m_1^2}{12   m_\gamma }
+\frac{m_\gamma}{12   }
+\frac{m_\gamma^2}{48  m_1 }
\right)
\right. \nonumber \\
&& \left.
+\frac{g^2 g'^2}{  (m_1+m_\gamma)^3}\left(
+\frac{9 }{32  }
+\frac{3 m_1^2}{32    m_\gamma^2 }
+\frac{9 m_1}{32     m_\gamma }
+\frac{3 m_\gamma}{32   m_1 }
\right)
\right. \nonumber \\
&& \left.
+\frac{g^2 g'^2}{ \left(m_1+\xi^{1/2}
  m_\gamma\right)^4}\left(
+\frac{29 \xi  m_1}{48   }
+\frac{5 m_1^3}{48   m_\gamma^2 }
+\frac{5 \xi^{1/2} m_1^2}{12   m_\gamma }
+\frac{\xi ^{3/2} m_\gamma}{3  }
\right)
\right. \nonumber \\
&& \left.
-\frac{g^2 g'^2 }{ \left(m_1+\xi^{1/2}
  m_\gamma\right)^3}\left(
-\frac{\xi }{4 }
-\frac{3 m_1^2}{32   m_\gamma^2 }
-\frac{9 \xi^{1/2} m_1}{32  m_\gamma }\right) \right)
\nonumber \\
&&+\frac{(eB)^2}{T^2} \frac{(g'^2-29 g^2) \zeta(3)}{1536\pi ^4}
\end{eqnarray}

\begin{eqnarray}
\label{piGCe}
 \Pi _e^{G^c}&=& \frac{\lambda^2 m_W^2}{g^2} \left[
-\frac{2   }{ \pi ^2}\ln\left(\frac{m_W}{T}\right)-\frac{4 T }{
\pi (m_1+ m_4)} \right.\nonumber \\
&& \left. +\frac{(eB)^2 T}{\pi  m_1 (m_1+m_4)^4}
\left(-\frac{1}{3  }
-\frac{4 m_4 }{3 m_1}-\frac{ m_4^2}{3 m_1^2 }
+\frac{3 (m_1+m_4)}{2  m_1 }
+\frac{ m_4 (m_1+m_4)}{2 m_1^2}\right)
+\frac{(eB)^2
 \zeta(5)}{128  \pi ^6 T^4} \right] \\ \nonumber \\
\label{piGCf}
 \Pi _f^{G^c}  &=& \frac{1}{v_0^2} \left[
\frac{T^2}{2} \left(m_b^2+m_t^2\right)
+ \frac{(eB)}{T} \frac{3(m_b^2-m_t^2)}{16 \pi  }
-\frac{3 m_b^2 m_t^2 }{\pi ^2 } \ln\left(\frac{4
  m_W}{T}\right)\right. \nonumber\\
&& \left. -\frac{(eB)^2}{T^2}\frac{133 (m_b^2+m_t^2)
\zeta(3)}{288 \pi ^4 }
+\frac{(eB)^2}{T^4}\frac{31 \zeta(5)}{256 \pi ^6 }
\left( m_b^4
+2m_b^2 m_t^2
 +4m_t^4\right)  \right] \\ \nonumber \\
\label{piGCg}
 \Pi _g^{G^c}&=& -\left(g^2-g'^2\right)\left(
\frac{ \xi ^2 m_W^2}{16 \pi ^2}\ln\left(\frac{m_W}{T}\right)
-\frac{ \xi ^2 T }{8\pi ( m_W+ m_Z)}
\right.\nonumber \\
&& + \left. \frac{(eB)^2 \xi ^2 T m_W^2}{\pi  m_W
  (m_W+m_Z)^4}
\left(\frac{1}{96}+\frac{ m_Z}{24 m_W}
+\frac{m_Z^2}{96 m_W^2}
-\frac{3  (m_W+m_Z) }{64 m_W
}-\frac{m_Z(m_W+m_Z)}{64  m_W^2
}\right)\right. \nonumber \\
&& \left.-\frac{(eB)^2\xi ^2\zeta(5)}{4096\pi ^6 T^4}\right)
\end{eqnarray}

\section*{Appendix B: Gauge boson self-energies in arbitrary gauge}

In this appendix, we list the results for the
self-energy diagrams depicted in Fig.~\ref{fig2}.

\subsection*{Photon ($\gamma$) self-energy in arbitrary gauge}

\begin{eqnarray}
\label{piGammaa}
 \Pi _a^\gamma  &=& \frac{g^2 g'^2}{\left(g^2+g'^2\right)} \left[ T^2 \left(
   \frac{\xi + 1}{4} - (\xi ^{3/2} +2) \frac{ m_W}{2 \pi T } \right)
 +\frac{ (eB)^2}{ 16 \pi m_W^2} \left(
   \left(\frac{1}{\xi ^{1/2}}-2 \right) \frac{T}{3 m_W} +\frac{7 \zeta(3) m_W^2}{4
 \pi ^3 T^2}\right) \right] \\ \nonumber \\
\label{piGammab}
 \Pi _b^\gamma &=& -\frac{g^2 g'^2  m_W^2}{\left(g^2+g'^2\right)}\left[
\frac{1}{4 \pi ^2}\ln\left(\frac{m_W}{T}\right)+\frac{2 T }{4\pi  (
  m_1+ m_W)} \right.\nonumber \\
&&-
\frac{(eB)^2T}{4\pi(m_1+m_W)^2} \left(\frac{1}{  m_W^3 }
+\frac{1}{4   (m_1+m_W)} \left(
\frac{3}{  m_1^2 }
-\frac{7 m_1}{  m_W^3}
-\frac{21}{ m_W^2 }
+\frac{m_W}{ m_1^3 } \right) -\frac{(m_1+m_W)^2}{ \xi^{1/2} m_W^3
  \left(m_1+\xi^{1/2} m_W\right)^2}\right. \nonumber \\
&& \left. \left. -\frac{ 1}{6 (m_1+m_W)^2}\left(
\frac{ 1}{ m_1 }
+\frac{1}{  m_W }
+\frac{m_1^2 }{ m_W^3 }
+\frac{m_W^2}{  m_1^3 }
+\frac{4 m_1}{  m_W^2 }
+\frac{4 m_W}{  m_1^2}
 \right)
\right)
+\frac{7 (eB)^2  \zeta(5)}{512  \pi ^6 T^4} \right] \\ \nonumber \\
\label{piGammac}
 \Pi _c^\gamma  &=& \frac{g^2 g'^2 T^2}{2 \left(g^2+g'^2\right)} \left[
\frac{1}{3} -\frac{m_1}{\pi T }\right] -\frac{ g^2 g'^2 (eB)^2}{48
   \left(g^2+g'^2\right) \pi  m_1^2} \left[ \frac{T}{m_1} +
   \frac{\zeta(3) m_1^2}{4 \pi ^3 T^2}\right] \\ \nonumber \\
\label{piGammad}
 \Pi _d^\gamma &=&
\frac{g^2 g'^2}{\left(g^2+g'^2\right)} \Biggl[ \frac{(2-\xi) T^2}{4}
+\frac{\xi
  Tm_W}{2\pi\left(1+\xi^{1/2}\right)}
\left(1 + \xi ^{1/2} +\xi \right) \nonumber \\
&& + \frac{(eB)^2 T}{2 \pi m_W^3}
\left( \frac{1}{4   \left(1+\xi^{1/2}\right)^2 } \left(
-\frac{37 }{2}\left(1+\xi^{1/2}\right)^2  + \frac{7
}{\left(1+\xi^{1/2}\right)^2}
+\frac{13}{2  \left(1+\xi^{1/2}\right) }
+\frac{14}{3   } \right) \right. \nonumber \\
&& \left. +\frac{\xi^{1/2}}{   \left(1+\xi^{1/2}\right)^2 }
\left(-\frac{15}{8}\left(1+\xi^{1/2}\right)^2
+\frac{49}{12  \left(1+\xi^{1/2}\right)^2 }
+\frac{27}{4   \left(1+\xi^{1/2}\right)}
+\frac{13}{3 } \right) \right. \nonumber \\
&& \left. +\frac{\xi }{   \left(1+\xi^{1/2}\right)^2 } \left(
\frac{49}{12  \left(1+\xi^{1/2}\right)^2}
+\frac{53}{8  \left(1+\xi^{1/2}\right)}
+ 5 \right) + \frac{\xi ^{3/2}}{2   \left(1+\xi^{1/2}\right)^2}
\left(\frac{7 }{2  \left(1+\xi^{1/2}\right)^2}
+5 \right) \right)\nonumber \\
&&-\frac{157(eB)^2  \zeta(3)}{384  \pi ^4 T^2} \Biggr]
\end{eqnarray}

\begin{eqnarray}
\label{piGammae}
 \Pi _e^\gamma &=& \frac{g^2 g'^2}{\left(g^2+g'^2\right)} \left(\frac{T^2}{6}
+\frac{(eB)^2 \zeta(3)}{64 \pi ^4 T^2}\right) \\ \nonumber \\
\label{piGammaf}
\Pi _f^\gamma  &=& \frac{g^2 g'^2}{\left(g^2+g'^2\right)}\left(
\frac{4 T^2}{9}-\frac{2 m_t^2 }{3  \pi ^2}\ln\left(\frac{4
m_W}{T}\right)+\frac{7 (eB)^2 \zeta(3)}{81
\pi ^4 T^2}-\frac{31 (eB)^2 m_t^2 \zeta(5)}{108 \pi ^6 T^4}
\right) \\ \nonumber \\
\label{piGammag}
\Pi _g^\gamma  &=& -\frac{g^2 g'^2}{\left(g^2+g'^2\right)}\left(
\frac{T^2}{12}+\frac{(eB)^2 \zeta(3)}{128 \pi ^4 T^2} \right)
\end{eqnarray}

\subsection*{Neutral gauge boson ($Z$) self-energy in arbitrary gauge}

\begin{eqnarray}
\label{piZa}
\Pi_a^Z
&=&\left(\frac{g^4}{g^2+g'^2}\right)\left\{
\frac{\left(1+\xi\right)}{4}T^2
-\left(\frac{4+2\xi ^{3/2}}{4\pi}\right)Tm_W
+\frac{(eB)^2}{16\pi m_W^2}
\left[\left(\frac{23-2\xi^{1/2}}{3\xi^{1/2}}\right)\frac{T}{m_W}
+\frac{7\zeta(3)}{4\pi^3}\frac{m_W^2}{T^2}\right]\right\} \\ \nonumber \\
\label{piZb}
\Pi _b^Z &=&
-\frac{\left(g^6+3 g^4 g'^2+5 g^2 g'^4+g'^6\right)m_W^2}
{8 g^2 \left(g^2+g'^2\right) \pi ^2}\ln\left(\frac{m_W}{T}\right)
-\frac{Tm_W^2}{4\pi(m_4+m_Z)}\left(\frac{2g'^4}
{\left(g^2+g'^2\right)}
\frac{(m_4+m_Z)}{(m_1+m_W)} \right. \nonumber \\
&& \left.+\frac{(g^2+g'^2)^2}{g^2}\right)
 +\frac{g'^4m_W^2 (eB)^2}{(g^2+g'^2)(m_1+m_W)^4}
\left[\frac{T}{48\pi m_1}
\left(7 -\frac{m_1^3}{m_W^3}
-68\frac{m_1^2}{m_W^2}
-53\frac{m_1}{m_W}
+ 4\frac{m_W}{m_1}+\frac{m_W^2}{m_1^2} \right. \right.
\nonumber\\
&&\left. \left.
-\frac{12}{\xi^{1/2}}\frac{m_1(m_1+m_W)^4}{m_W^3
  \left(m_1+\xi^{1/2}
m_W\right)^2}\right)
- \frac{7\zeta(5)}{512\pi ^6}
\frac{(m_1+m_W)^4}{T^4}\right] \\ \nonumber \\
\label{piZc}
\Pi_c^Z &=&
\frac{g^4+g'^4}{g^2+g'^2}
\frac{T^2}{12}  -\left[\frac{2(g^2-g'^2)^2m_1
+(g^2+g'^2)^2(m_3+m_4)}{g^2+g'^2}\right]\frac{T}{16\pi}
- \frac{(g^2-g'^2)^2}{g^2+g'^2}
\frac{(eB)^2}{192\pi m_1^2}\left[\frac{T}{m_1}
+\frac{\zeta(3)}{4\pi^3}\frac{m_1^2}{T^2}\right] \nonumber \\ \\
\label{piZd}
\Pi _d^Z &=&
\frac{g^4}{g^2+g'^2}\left\{
\left(1-\frac{\xi}{2}\right)\frac{T^2}{2}
+\frac{\xi\left(1+\xi^{1/2}+\xi\right)}{2\pi(1+\xi^{1/2})}Tm_W
+\frac{1}{16\pi}\left[ \frac{14}{\left(1+\xi^{1/2}\right)^4}
\left( 1 + \frac{7}{3}\xi^{1/2}
+ \frac{7}{3}\xi + \xi^{3/2}\right)\right.\right.
\nonumber \\
&&+\frac{1}{\left(1+\xi^{1/2}\right)^3}
\left( 13 + 54\xi^{1/2} + 53\xi \right)
+\frac{8}{\left(1+\xi^{1/2}\right)^2}
\left( \frac{7}{6} + \frac{13\xi^{1/2}}{3}
+5\xi +\frac{5\xi^{3/2}}{2} \right) \nonumber \\
&&-\left.\left.\left( 37 + 15\xi^{1/2} \right)
-\frac{157\zeta(3)}{24\pi^3}\frac{m_W^3}{T^3}
\right]\frac{(eB)^2T}{m_W^3}\right\} \\ \nonumber \\
\label{piZe}
\Pi _e^Z &=&
\left(\frac{g^4+g'^4}{g^2+g'^2}\right) \frac{T^2}{12}
+\frac{(g^2-g'^2)^2}{g^2+g'^2}\frac{\zeta(3)}{256\pi ^4}
\frac{(eB)^2}{T^2} \\ \nonumber \\
\label{piZf}
\Pi _f^Z &=&
\frac{g'^2 \left(3 g^2-g'^2\right) }{6 \left(g^2+g'^2\right)
\pi ^2} m_t^2 \ln\left(\frac{4m_W}{T}\right)
\nonumber\\
&& +
\left(-\frac{g'^2}{3}+\frac{4 g'^4}{9 \left(g^2+g'^2\right)}+\frac{1}{8}
\left(g^2+g'^2\right)\right)
\left\{ T^2 - \frac{(eB)^2}{T^2}
\left[\frac{31\zeta(5)}{48\pi ^6}\frac{m_t^2}{T^2}
-\frac{7\zeta(3)}{36\pi ^4}\right] \right\} \\ \nonumber \\
\label{piZg}
\Pi _g^Z &=&
-\frac{g^4}{\left(g^2+g'^2\right)}
\left[\frac{T^2}{12}+\frac{\zeta(3)}{128\pi ^4}
\frac{(eB)^2}{T^2}\right]
\end{eqnarray}

\subsection*{Charged gauge boson ($W$) self-energy in arbitrary gauge}
\small

\begin{eqnarray}
\label{piWa}
 \Pi _a^W &=& 2 g^2 T^2 (\xi + 1)
- \frac{(2+\xi ^{3/2})g^2 T}{4\pi\left(g^2+g'^2\right)} \left(
\left(g^2+g'^2\right) m_W
+g^2 m_Z
+g'^2 m_\gamma
\right)
\nonumber \\
&& -\frac{g^2 (eB)^2}{128 \pi m_W^2} \left( \frac{4 T}{3 m_W}
\left(2-\frac{23 }{\xi^{1/2} }\right)
+\frac{7 \zeta(3)m_W^2}{\pi ^3 T^2} \right) \\ \nonumber \\
\label{piWb}
 \Pi _b^W &=&
-\frac{\left(g^2+g'^2\right)  m_W^2}{8 \pi ^2} \ln
\left(\frac{m_W}{T}\right)
-\frac{T m_W^2}{4 \pi \left(g^2+g'^2\right) } \left(
\frac{g^2 \left(g^2+g'^2\right) }{m_4+ m_W}
+\frac{g'^4 }{m_1+m_Z}
+\frac{g^2 g'^2}{ m_1+m_\gamma}\right) \nonumber \\
&& +(eB)^2 T
\left(
-\frac{g^2 m_W}{48 \pi   (m_4+m_W)^4}\left(
\frac{ m_4^2 }{   m_W^2 }
+\frac{4  m_4 }{   m_W }
+1
\right)
\right. \nonumber \\
&& \left.
-\frac{7  g^2  }{32 \pi
  (m_4+m_W)^3}\left(
\frac{ m_4 }{ m_W }
+3
\right)
+\frac{g^2 }{8 \pi  m_W}\left(
\frac{1}{ (m_4+m_W)^2}
-\frac{1}{\xi^{1/2} \left(m_4+\xi^{1/2} m_W\right)^2}
\right)
\right. \nonumber \\
&& \left.
-\frac{g'^4 m_W^2}{48\pi  m_1 \left(g^2+g'^2\right)
  (m_1+m_Z)^4}\left(
1
+\frac{4 m_Z}{m_1}
+\frac{m_Z^2}{m_1^2}
\right)
+\frac{g'^4 m_W^2}{32 \left(g^2+g'^2\right)\pi  m_1^2
  (m_1+m_Z)^3}\left(
3
+\frac{m_Z}{m_1}
\right)
\right. \nonumber \\
&& \left.
-\frac{g^2 g'^2 m_W^2}{48 \left(g^2+g'^2\right)\pi  m_1
  (m_1+m_\gamma)^4}\left(
1
+\frac{4 m_\gamma}{m_1}
+\frac{m_\gamma^2}{m_1^2}
\right)
+\frac{g^2 g'^2 m_W^2}{32 \left(g^2+g'^2\right) \pi  m_1^2
  (m_1+m_\gamma)^3}\left(
3
+\frac{m_\gamma}{m_1}\right)
\right)
\nonumber \\
&& +\frac{(eB)^2 m_W^2 \zeta(5)}{2048 \pi ^6 T^4}
\left(g'^4-15 g^4 - 14 g^2 g'^2  \right) \\ \nonumber \\
\label{piWc}
 \Pi _c^W &=&
\frac{g^2 T^2}{12}
-\frac{g^2 T}{16 \pi}
\left(2 m_1
+m_3
+m_4\right)
- \frac{g^2 (eB)^2}{192 \pi m_1^2} \left(
\frac{ T}{m_1}
+\frac{ \zeta(3)m_1^2}{4 \pi ^3 T^2}
\right) \\ \nonumber \\
\label{piWd}
 \Pi _d^W &=&
\frac{T^2 g^2}{4} \left( 2- \xi \right)
+\frac{T}{4 \pi \left(g^2+g'^2\right)} \left(
-\frac{g^4}{(  m_W+    m_Z)}\left(
\frac{m_W^4}{m_Z^2 }
+\frac{ m_W^3}{  m_Z}
+\frac{ m_Z^3}{  m_W }
+\frac{m_Z^4}{  m_W^2} \right)
\right. \nonumber \\
&& \left.
+\frac{g^4}{  \left(\xi  m_W^2-m_Z^2\right)}\left(
\xi ^{5/2}(m_W^3-m_Z^3)
+\frac{ m_W^5}{  m_Z^2 }
-\frac{ m_Z^5}{ m_W^2 }
 \right)
-\frac{g^2 g'^2 }{ (    m_W+    m_\gamma)}\left(
\frac{m_W^4}{ m_\gamma^2 }
+\frac{ m_W^3}{ m_\gamma }
+\frac{ m_\gamma^3}{ m_W }
+\frac{ m_\gamma^4}{ m_W^2 } \right)
\right. \nonumber \\
&& \left.
+\frac{g^2 g'^2 }{  \left(\xi  m_W^2-m_\gamma^2\right)}\left(
\xi ^{5/2} m_W^3
-\frac{ m_\gamma^5}{m_W^2} \right)
+\frac{g^2 g'^2}{\left(\xi  m_\gamma^2-m_W^2\right)} \left(
\xi ^{5/2} m_{\gamma}^3
-\frac{m_W^5}{m_\gamma^2 } \right)
\right) \nonumber \\
&& +\frac{(eB)^2 T }{\pi \left(g^2+g'^2\right)} \left(
-\frac{g^4}{(m_W+m_Z)^4}\left(
\frac{3 m_W}{4    }
+\frac{5  m_W^3}{48 m_Z^2 }
+\frac{5  m_W^2}{12 m_Z }
+\frac{11 m_Z}{12}
+\frac{13 m_Z^2}{16 m_W }
+\frac{ m_Z^3}{3 m_W^2 }
\right)
\right. \nonumber \\
&& \left.
-\frac{ g^4}{(m_W+m_Z)^3}\left(
\frac{185 }{96}
+\frac{21  m_W^2}{32 m_Z^2 }
+\frac{63  m_W}{32 m_Z }
+\frac{17 m_Z}{32 m_W }
-\frac{ m_Z^2}{4 m_W^2 }
\right)
\right. \nonumber \\
&& \left.
-\frac{g^4}{(m_W+m_Z)^2}\left(
\frac{1}{3 m_W }
+\frac{3  m_W}{8 m_Z^2 }
+\frac{3 }{4 m_Z }
-\frac{ m_Z}{12 m_W^2 }
\right)
+\frac{ g^4 }{ \left(m_W+\xi^{1/2} m_Z\right)^3}\left(
\frac{7  \xi }{4 }
+\frac{21  m_W^2}{32m_Z^2 }
+\frac{63  \xi^{1/2} m_W}{32m_Z }
\right)
\right. \nonumber \\
&& \left.
+\frac{g^4}{\left(\xi^{1/2} m_W+m_Z\right)^4}\left(
\frac{5 \xi ^{3/2} m_W}{48}
+\frac{5  \xi  m_Z}{12}
+\frac{29 \xi^{1/2} m_Z^2}{48 m_W }
+\frac{ m_Z^3}{3 m_W^2 }
\right)
\right. \nonumber \\
&& \left.
-\frac{g^4 }{ \left(\xi^{1/2} m_W+m_Z\right)^3}\left(
\frac{3  \xi }{32 }
+\frac{9  \xi^{1/2} m_Z}{32 m_W}
+\frac{ m_Z^2}{4 m_W^2 }
\right)
+\frac{g^4 }{\left(\xi^{1/2} m_W+m_Z\right)^2} \left(
\frac{ \xi^{1/2}}{3m_W }
+\frac{3  \xi ^{3/2} m_W}{8 m_Z^2 }
+\frac{3  \xi }{4 m_Z }
-\frac{ m_Z}{12 m_W^2 }
\right)
\right. \nonumber \\
&& \left.
+\frac{g^4 }{\left(m_W+\xi^{1/2} m_Z\right)^4}\left(
\frac{29  \xi  m_W}{48 }
+\frac{5  m_W^3}{48 m_Z^2 }
+\frac{5  \xi^{1/2} m_W^2}{12m_Z }
+\frac{ \xi ^{3/2} m_Z}{3 }
\right)
\right. \nonumber \\
&& \left.
+\frac{g^4  }{ \left(m_W+\xi^{1/2}m_Z\right)^2}\left(
\frac{ \xi }{2 m_W }
+\frac{3 m_W}{8 m_Z^2 }
+\frac{3\xi^{1/2}}{4 m_Z }
+\frac{\xi ^{3/2} m_Z}{4 m_W^2 }
\right)
\right. \nonumber \\
&& \left.
-\frac{g^4}{\left(\xi^{1/2} m_W+\xi^{1/2} m_Z\right)^2}\left(
\frac{\xi ^{3/2}}{2 m_W }
+\frac{3\xi ^{3/2} m_W}{8 m_Z^2 }
+\frac{3  \xi ^{3/2}}{4 m_Z }
+\frac{\xi ^{3/2} m_Z}{4 m_W^2 }
\right)
\right. \nonumber \\
&& \left.
-\frac{ g^2 g'^2}{(m_W+m_\gamma)^4}\left(
\frac{3  m_W}{4}
+\frac{5  m_W^3}{48 m_{\gamma}^2}
+\frac{5  m_W^2}{12m_\gamma }
+\frac{11  m_\gamma}{12}
+\frac{13  m_\gamma^2}{16m_W}
+\frac{ m_\gamma^3}{3m_W^2 }
\right)
\right. \nonumber \\
&& \left.
-\frac{ g^2 g'^2}{(m_W+m_\gamma)^3}\left(
\frac{185 }{96}
+\frac{21  m_W^2}{32m_{\gamma}^2}
+\frac{63  m_W}{32 m_\gamma }
+\frac{17  m_\gamma}{32m_W}
-\frac{ m_\gamma^2}{4m_W^2 }
\right)
\right. \nonumber
\end{eqnarray}
\normalsize

\begin{eqnarray}
&& \left.
-\frac{g^2 g'^2}{(m_W+m_\gamma)^2}\left(
\frac{1}{3m_W}
+\frac{3  m_W}{8 m_{\gamma}^2}
+\frac{3 }{4m_{\gamma} }
-\frac{ m_\gamma}{12 m_W^2 }
\right)
\right. \nonumber \\
&& \left.
+\frac{g^2 g'^2 }{\left(\xi^{1/2} m_W+m_\gamma\right)^4}\left(
\frac{5  \xi ^{3/2} m_W}{48}
+\frac{5  \xi  m_\gamma}{12 }
+\frac{29 \xi^{1/2} m_\gamma^2}{48 m_W }
+\frac{ m_\gamma^3}{3 m_W^2 }
\right)
\right. \nonumber \\
&& \left.
+\frac{g^2 g'^2 }{\left(\xi^{1/2} m_W+m_\gamma\right)^2}\left(
\frac{ \xi^{1/2}}{3m_W}
+\frac{3  \xi ^{3/2} m_W}{8 m_\gamma^2 }
+\frac{3  \xi }{4 m_\gamma }
-\frac{ m_\gamma}{12 m_W^2 }
\right)
\right. \nonumber \\
&& \left.
+\frac{ g^2 g'^2 }{ \left(m_W+\xi^{1/2} m_\gamma\right)^4}
\left(
\frac{29  \xi  m_W}{48 }
+\frac{5  m_W^3}{48 m_\gamma^2 }
+\frac{5  \xi^{1/2} m_W^2}{12m_{\gamma} }
+\frac{ \xi ^{3/2} m_\gamma}{3 }
\right)
\right. \nonumber \\
&& \left.
-\frac{g^2 g'^2  }{\left(\xi^{1/2}m_W+m_\gamma\right)^3}\left(
\frac{3 \xi }{32}
+\frac{9 \xi^{1/2} m_\gamma}{32 m_W }
+\frac{ m_\gamma^2}{4 m_W^2 }
\right)
+\frac{ g^2 g'^2  }{\left(m_W+\xi^{1/2} m_\gamma\right)^3}\left(
\frac{7  \xi }{4 }
+\frac{21  m_W^2}{32 m_\gamma^2 }
+\frac{63  \xi^{1/2} m_W}{32 m_\gamma }
\right)
\right. \nonumber \\
&& \left.
+\frac{g^2 g'^2  }{\left(m_W+\xi^{1/2} m_\gamma\right)^2}\left(
\frac{\xi }{2 m_W }
+\frac{3  m_W}{8 m_\gamma^2 }
+\frac{3  \xi^{1/2}}{4 m_\gamma }
+\frac{ \xi ^{3/2} m_\gamma}{4 m_W^2}
\right)
\right. \nonumber \\
&& \left.
-\frac{g^2 g'^2 \xi ^{1/2}}{ \left( m_W+
  m_\gamma\right)^2}\left(
\frac{1 }{2  m_W }
+\frac{3  m_W}{8 m_\gamma^2 }
+\frac{3 }{4 m_\gamma }
+\frac{ m_\gamma}{4 m_W^2}\right)
\right)
-\frac{109 (eB)^2 \zeta(3) g^2}{768 T^2 \pi ^4 } \\ \nonumber \\
\label{piWe}
\Pi _e^W &=&
\frac{g^2 T^2}{12}+\frac{(eB)^2 g^2 \zeta(3)}{256 \pi ^4 T^2} \\ \nonumber \\
\label{piWf}
 \Pi _f^W &=&
\frac{g^2 T^2}{4}   +  \frac{(eB)^2 g^2 }{\pi ^4 T^2}
\left(
\frac{133 \zeta(3)}{2304 }
-\frac{31 \zeta(5)\left(m_b^2 + 4 m_t^2 \right)}{1536 \pi ^2 T^2}
\right) \\ \nonumber \\
\label{piWg}
 \Pi _g^W&=&
-\frac{g^2 T^2}{12}-\frac{g^2 (eB)^2 \zeta(3)}{256 \pi ^4 T^2}
\end{eqnarray}

\section*{Appendix C: Ghosts self-energies in arbitrary gauge}

In this appendix, we list the results for the
self-energy diagrams depicted in Fig.~\ref{fig6}.

\subsection*{$\gamma$ boson associated ghost ($\eta_\gamma$) self-energy in arbitrary gauge}

\begin{eqnarray}
\label{piGhGhAa}
\Pi _a^{\eta_\gamma} = \Pi _b^{\eta_\gamma} &=& 0
\end{eqnarray}

\subsection*{$Z$ boson associated ghost ($\eta_Z$) self-energy in arbitrary gauge}

\begin{eqnarray}
\label{piGhGhZa}
\Pi _a^{\eta_Z} &=& 0 \\ ~ \nonumber  \\
\label{piGhGhZb}
\Pi _b^{\eta_Z}&=& \frac{g(g^2-g'^2)}{(g^2+g'^2)^{1/2}} \frac{\xi^2 T m_Z}{8\pi} \Biggl(
\frac{m_W}{m_1 + \xi^{1/2}m_W}
- \frac{(g^2+g'^2)^{3/2}}{g(g^2-g'^2)}\frac{m_Z}{2(m_4 + \xi^{1/2}m_Z)} \nonumber \\
&&- (eB)^2 \frac{\xi^{1/2}(m_1^2 + \xi^{1/2}m_1 m_W+ \xi m_W^2)}{96 m_1^3 m_W^2(m_1 + \xi^{1/2}m_W)}
\Biggr)
\end{eqnarray}

\subsection*{$W$ boson associated ghost ($\eta_W$) self-energy in arbitrary gauge}

\begin{eqnarray}
\label{piGhGhWa}
\Pi _a^{\eta_W} &=& 0 \\ ~ \nonumber  \\
\label{piGhGhWb}
\Pi _b^{\eta_W} &=& \frac{g^2\xi^2 T m_W}{16\pi} \Biggl(
\frac{m_W}{m_3+ \xi^{1/2}m_W}
- \frac{m_W}{m_4+ \xi^{1/2}m_W} + \frac{g^2-g'^2}{g(g^2+g'^2)^{1/2}}\frac{m_Z}{m_1+ \xi^{1/2}m_Z}
\Biggr) \nonumber \\
&&+ (eB)^2\frac{g^2\xi^{1/2} T m_W}{384\pi} \Biggl(
\frac{(m_4^2+4\xi^{1/2}m_4 m_W + 7\xi m_W^2)}{m_W^2(m_4 + \xi^{1/2}m_W)^4}
- \frac{(g^2-g'^2)}{g(g^2+g'^2)^{1/2}} \frac{\xi^{3/2}m_Z(7m_1^2 + 4\xi^{1/2}m_1 m_Z + \xi m_Z^2)}{m_1^3(m_1 + \xi^{1/2}m_Z)^4} \nonumber \\
&&+ \frac{2(m_3^2 +4\xi^{1/2}m_3 m_W+ \xi m_W^2)}{m_W^2(m_3 + \xi^{1/2}m_W)^4}
- \frac{3(m_3 +3\xi^{1/2}m_W)}{m_W^2(m_3 + \xi^{1/2}m_W)^3}
\Biggr)
\end{eqnarray}

\end{widetext}

\end{document}